\documentclass[journal]{IEEEtran}
\usepackage{amsmath,amssymb,amsfonts}
\usepackage[caption=false,font=normalsize,labelfont=sf,textfont=sf]{subfig}
\usepackage{stfloats}
\usepackage{verbatim}
\usepackage{graphicx}
\usepackage{bm}
\usepackage{cite}
\usepackage{color}
\usepackage{subfig}
\usepackage[]{hyperref}

\usepackage[linesnumbered, ruled]{algorithm2e}
\SetKwRepeat{Do}{do}{while}%
\begin{document}

\title{Time-Slotted Multi-Cluster UAV AirComp with Energy-Awareness: A Pointer Network-Assisted Soft Actor-Critic Learning Framework}
\author{Xunqiang Lan,~\IEEEmembership{Graduate Student Member,~IEEE,}
		Xiao Tang,~\IEEEmembership{Member,~IEEE,}
        Ruonan Zhang,~\IEEEmembership{Member,~IEEE,}\\
        Qinghe Du,~\IEEEmembership{Member,~IEEE,}
        and Tony Q.S. Quek,~\IEEEmembership{Fellow,~IEEE}

\thanks{X. Lan and R. Zhang are with the School of Electronics and Information, Northwestern Polytechnical University, Xi'an 710072, China. (email: lanxunqiang@mail.nwpu.edu.cn, rzhang@nwpu.edu.cn)}
\thanks{X. Tang and Q. Du are with the School of Information and Communication Engineering, Xi'an Jiaotong University, Xi'an 710049, China. (e-mail: tangxiao@xjtu.edu.cn, duqinghe@mail.xjtu.edu.cn)}
\thanks{T. Q.S. Quek is with the Information Systems Technology and Design Pillar, Singapore University of Technology and Design, Singapore 487372, Singapore. (e-mail: tonyquek@sutd.edu.sg)}
}

\maketitle

\begin{abstract}
Over-the-air computation (AirComp) has emerged as a promising approach for massive data aggregation, which is yet challenged by the channel variations, task distributions, and inherent energy limitation of the computation nodes. In this paper, we propose an unmanned aerial vehicle (UAV)-assisted Aircomp system to serve multi-cluster computation tasks over time, where the UAV mobility-facilitated spatial and time diversity is exploited for efficient and accurate data computation. Specifically, we aim for the minimization of AirComp aggregation error and the energy consumption by jointly optimizing the transceiver beamforming, normalizing factors, sensor scheduling, and UAV trajectory. To solve the formulated problem, we decompose it into two layers where the inner layer addresses the optimization-based AirComp transceiver design, and the outer layer focuses on the deep reinforcement learning (DRL)-based scheduling and trajectory design. In particular, a pointer network actor-critic learning is developed to tackle the binary scheduling problem, and a soft actor-critic DRL algorithm is employed to determine the UAV trajectory. Simulation results validate the convergence of the proposed hierarchical learning framework and demonstrate its significant performance gains in terms of AirComp aggregation error and energy consumption as compared with baseline schemes.
\end{abstract}


\section{Introduction}
\IEEEPARstart{W}{ith} the rapid advancement of unmanned aerial vehicle (UAV) technology and the evolution toward the sixth generation (6G), low-altitude wireless networks (LAWNs) have emerged as a key technology for future wireless networks~\cite{lawn1,lawn2}. As a crucial component of LAWNs, UAVs can be deployed as aerial base stations close to ground devices, providing enhanced line-of-sight (LoS) communications~\cite{uav1}. By leveraging the mobility and three-dimensional deployment capabilities, UAVs can not only expand network coverage but also support multi-cluster operations in LAWNs, enabling flexible and efficient networks for intelligent 6G applications~\cite{LAWUAV}. Meanwhile, with the continuous emergence of new intelligent architectures~\cite{Zhang2025Semantic}, including edge intelligence, distributed learning, and collaborative intelligence in LAWNs~\cite{lawn3}, the number and scale of intelligent devices have surged, making massive data transmission and computational processing increasingly challenging~\cite{6G}.

However, under traditional network architectures, a large amount of data generated by terminal devices need to be individually transmitted to the cloud for centralized computing~\cite{spectrum}. This ``transmit-then-compute" data processing protocol faces challenges of high latency and high energy consumption in large-scale LAWNs~\cite{latency}. The high energy consumption is particularly concerning for UAVs and sensors, which are constrained by strict size, weight, and battery capacity limitations. In this regard, over-the-air computation (AirComp) leverages the waveform or signal superposition characteristics of wireless multiple-access channels to achieve instantaneous aggregation of distributed data during transmission~\cite{aircomp}. This novel data processing paradigm effectively reduces the transmission latency, redundant data exchange operations, and spectrum occupation, thereby significantly mitigating data processing latency and energy consumption while improving spectrum resource utilization~\cite{ac0}.

Due to the low latency, energy efficiency, and high spectral efficiency, AirComp is particularly attractive for delay-critical applications that compute aggregate functions over a large amount of data instead of reconstructing individual data, such as consensus control, collaborative sensing, and distributed machine learning~\cite{app1,app2}. In particular, for large-scale and diversified scenarios, multi-cluster AirComp has emerged as an effective solution~\cite{uac3}. However, the widely adopted single-slot AirComp framework relies on instantaneous channel conditions, and thus, the performance is highly unpredictable. In practical scenarios, due to differences in channel gains among distributed devices, AirComp is highly sensitive to channel variations~\cite{csi}. In particular, some devices in deep fading channel states are unable to provide the power required for signal amplitude alignment, directly compromising the accuracy and reliability of computations. Towards this issue, multi-slot AirComp frameworks have been gradually introduced to compensate for the shortcomings of the single-slot framework. The core of the multi-slot AirComp is scheduling devices in batches according to instantaneous channel conditions and prioritizing transmissions from devices with better channels, thereby enabling accurate and efficient AirComp.

Although multi-slot AirComp frameworks can improve aggregation accuracy by actively exploiting the time diversity, the performance is still fundamentally limited by the channel quality. In this regard, UAV-assisted AirComp offers a promising solution by enhancing channel conditions and supporting more reliable transmission~\cite{Ge2022Dynamic}. However, existing studies on UAV-assisted AirComp primarily focus on performance optimization within single-scenario, single-slot AirComp systems, with insufficient exploration on UAV dynamics over time to support diversified computation tasks.
In UAV-assisted time-slotted multi-cluster AirComp, UAV mobility intensifies channel dynamics and introduces strong coupling among transmission, scheduling, and trajectory design. Conventional optimization approaches may face the challenges in handling such high-dimensional and temporally correlated decisions in real time. In this regard, deep reinforcement learning (DRL) enables the learning of adaptive policies through continuous interaction with the environment, making it well suited for sequential and dynamic control tasks~\cite{drlcom,drlcom1}. By integrating DRL with traditional optimization, long-term adaptive decisions such as cluster scheduling and trajectory planning can be managed by DRL, while analytical optimization addresses short-timescale variables such as power control and beamforming~\cite{drlaerial}. This integration facilitates the development of a learning framework that effectively balances adaptability and efficiency, achieving reliable and energy-efficient AirComp in LAWNs.

In this paper, we consider a UAV-assisted AirComp system to serve multi-cluster computation tasks over time. Accordingly, we propose a pointer network-assisted soft actor-critic (SAC) approach to jointly investigate the computation accuracy and energy consumption. The main contributions can be summarized as follows:
\begin{itemize}
\item We investigate a UAV-assisted multi-cluster multi-slot AirComp framework, where a UAV performs data aggregation over multiple sensor clusters. We jointly consider the multi-slot scheduling among clusters and the adaptive UAV trajectory management to exploit diversity gains for energy-efficient operations.
\item We formulate the problem to minimize the weighted sum of AirComp aggregation error and the system energy consumption, where the AirComp transceiver beamforming, time-slotted cluster scheduling, and UAV trajectory are jointly optimized.
\item We develop a hierarchical learning framework with a structured inner signal processing module and a learning-based outer decision module to solve coupled problems. The inner AirComp transceiver optimization is solved by alternating optimization, while the outer layer uses a pointer network and the SAC algorithm to jointly perform scheduling and UAV trajectory control.
\item Simulation results validate the convergence of the proposed hierarchical learning framework, and show that it achieves more adaptive device scheduling, higher AirComp accuracy, and enhanced energy efficiency as compared to baseline schemes.
\end{itemize}

The rest of this paper is organized as follows. In Sec.~\ref{sec:rw}, we review the related works. In Sec.~\ref{sec:sys}, we introduce the system model, as well as the formulation and analysis of the problem. In Sec.~\ref{sec:inner}, the inner problem is solved for the AirComp transceiving strategy optimization. In Sec.~\ref{sec:outer}, a pointer network-assisted SAC algorithm for solving the sensor scheduling and UAV trajectory is introduced. Sec.~\ref{sec:sim} provides the simulation results to demonstrate the performance, and finally, Sec.~\ref{sec:con} concludes this paper.

\section{Related Works} \label{sec:rw}
\subsection{UAV-Assisted AirComp}
UAV-facilitated wireless communication naturally appears as a flexible and effective solution for AirComp, and therefore attracted research interests in UAV-assisted AirComp networks~\cite{uac0}. In~\cite{uac2}, the authors considered a UAV-aided AirComp system, jointly optimizing trajectory, receiver normalization factors, and transmit power of sensors, and proposed a novel equivalent problem transformation to minimize the mean-square error (MSE) of AirComp. In~\cite{uac4}, the authors investigated the impact of imperfect channel state information on the MSE performance of AirComp, demonstrating the significance of accurate channel estimation for AirComp. In~\cite{uac5}, a non-coherent UAV-assisted AirComp scheme was proposed, where amplitude modulation of complementary sequences enables reliable majority voting in fading channels. In~\cite{uac6}, a multi-UAV data collection framework was developed using space-timeline coding to exploit full spatial diversity, thereby reducing operation time and AirComp MSE.

Moreover, since both UAVs and distributed devices are subject to stringent energy constraints, energy efficiency has become a critical concern in UAV-assisted AirComp systems. In~\cite{uac1}, the authors investigated a power-limited multi-slot UAV-assisted AirComp framework, jointly optimizing resource allocation and UAV trajectory to minimize both the MSE of aggregation and the energy consumption of UAV flight. However, research on energy-efficient design in UAV-assisted AirComp remains limited. This lack of systematic investigation into energy efficiency poses a significant barrier to the practical deployment and large-scale implementation of UAV-assisted AirComp.

\subsection{Learning-Based AirComp}
With the rapid advancement of machine learning techniques, learning-based AirComp has emerged as a promising paradigm to overcome the limitations of conventional model-driven approaches, such as high computational complexity, strong dependence on precise channel models, and poor adaptability to time-varying and large-scale networks. By enabling adaptive resource allocation under time-varying channels and large-scale networks, learning-based methods effectively reduce aggregation errors and improve system efficiency. In~\cite{ddrl}, a dual time-scale deep reinforcement learning (DRL) framework was proposed for joint 3-D UAV deployment and dynamic service optimization in UAV-assisted intelligent transportation systems, achieving significant improvements in coverage and resource utilization. In~\cite{ac2}, an over-the-air federated edge learning scheme was developed to adapt transmission power for mitigating aggregation errors and enhancing communication efficiency. In~\cite{co}, an SAC-based reinforcement learning approach was introduced for coexisting AirComp and communication systems, jointly optimizing transmission, scheduling, and UAV trajectory to maximize user throughput while maintaining computation accuracy.

Building on the learning-driven frameworks, some studies have integrated AirComp with emerging technologies to enhance performance and adaptability. In~\cite{ac4}, a reconfigurable intelligent surface (RIS)-assisted AirComp was investigated, where transceiver parameters and RIS phase shifts were jointly optimized to improve signal alignment under fading conditions. In~\cite{ac5}, RIS and wireless power transfer were combined to improve energy efficiency and reduce AirComp distortion. In~\cite{unfold}, multi-cluster AirComp was explored by modeling inter-cluster interference as a graph, and an unfolding-based learning architecture was proposed to jointly optimize transmission scalars and receive beamforming for maximizing the weighted-sum AirComp rate. However, UAV-assisted time-slotted and multi-cluster AirComp systems under energy constraints remain underexplored, calling for adaptive learning frameworks to enable scalable and efficient operations.

\begin{figure*}[t]
	\centering
	\includegraphics[width=0.98\linewidth]{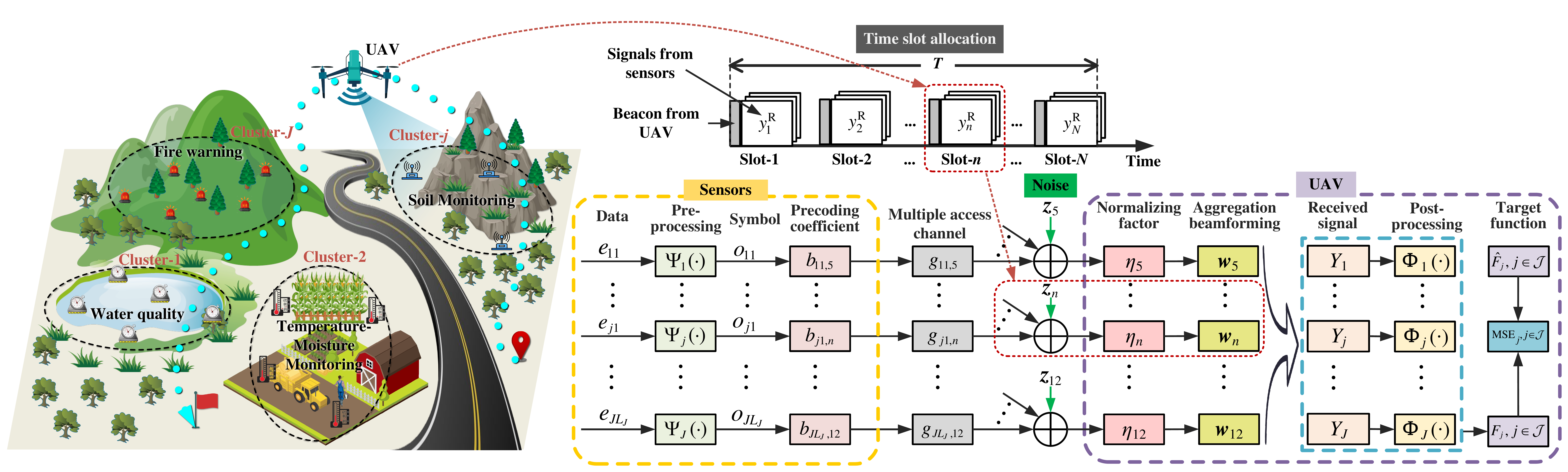}
	\vspace{-2.5mm}
	\caption{System model.}
	\vspace{-3.5mm}
	\label{fig:sys}
\end{figure*}

\section{System Model and Problem Formulation} \label{sec:sys}
\subsection{Network Model}
As shown in Fig.~\ref{fig:sys}, we consider a UAV-assisted multi-cluster AirComp network with $L$ sensors, denoted as $\mathcal{L} = \left\{1,2, \cdots, L\right\}$. The sensors are deployed to gather local data (e.g., temperature, pressure, power, or humidity) for specific tasks, and are divided into $J$ clusters based on task types, denoted as $\mathcal{J} = \left\{1,2, \cdots, J\right\}$. For the $j$-th cluster, there are $L_j$ sensors, denoted as $\mathcal{L}_j = \left\{1,2, \cdots, L_j\right\}$ with $\mathcal{L}_j\cap \mathcal{L}_{l}=\emptyset, \forall j\neq l$, $j,l\in \mathcal{J}$, and $\mathcal{L}\triangleq\cup_{j\in\mathcal{J}}\mathcal{L}_j$. For the $i$-th sensor in the $j$-th cluster with $i\in \mathcal{L}_j$ and $j\in \mathcal{J}$, its coordinates are denoted by $\bm{q}_{ji}=[x_{ji}, y_{ji}, 0]$. To facilitate fast wireless data aggregation among multiple sensors within each cluster, we deploy a rotary-wing UAV as an aerial aggregation center for data aggregation. The sensors has one single antenna, while the UAV is equipped with $K$ antennas to enhance the reception.

Specifically, we expatiate this process with a time-slotted model, where the UAV flight time $T$ is discretized into $N$ slots, denoted by $\mathcal{N} = \left\{1,2,\cdots,N\right\}$, each with a duration of $\delta = T/N$. In this regard, discrete waypoints serve as approximations for traversing corresponding time slots. The UAV's discrete waypoints are denoted by $\bm{q}^\text{u}=\left\{\bm{q}^\text{u}_n\right\}_{n=0,1,\cdots,N}$, where $\bm{q}^\text{u}_n=[x^\text{u}_n, y^\text{u}_n, H_n]$ is the three-dimensional coordinates of the UAV's discrete waypoint in the $n$-th slot, with $H_n$, $x^\text{u}_n$ and $y^\text{u}_n$ being the flight altitude, $\textit{x}$- and $\textit{y}$-coordinates of UAV, respectively. In addition, the UAV initiates the mission from a predefined starting point $\bm{q}_o$ and ends at a designated destination $\bm{q}_f$. The mobility constraints of the UAV are defined by
\begin{equation}
	\bm{q}^\text{u}_0 = \bm{q}_o,~\bm{q}^\text{u}_{N} = \bm{q}_f,
\end{equation}
\begin{equation}
	H_{\min} \leq H_n \leq H_{\max},~\forall n\in\mathcal{N},
\end{equation}
where $H_{\min}$ and $H_{\max}$ are the minimum and maximum flight heights of the UAV, respectively.

In each slot, sensors communicate with the UAV, and the channels from sensors to UAV follow the probabilistic LoS model~\cite{los}. We assume that the UAV can acquire perfect channel state information through control commands and pilot signals. Specifically, the channel (in dB) from the $i$-th sensor in the $j$-th cluster to the UAV in slot $n$ is given by
\begin{equation} \label{eq:gain}
	\begin{split}
		G_{ji,n} =\:&20\log_{10}\left(\frac{4\pi f_c}{c}\|\bm{q}^\text{u}_n-\bm{q}_{ji}\|\right)\\
		 &+ \widetilde{P}^{\mathsf{LoS}}_{ji,n} \mu_{\mathsf{LoS}} + \left(1-\widetilde{P}^{\mathsf{LoS}}_{ji,n}\right) \mu_{\mathsf{NLoS}},~\forall n\in\mathcal{N},
	\end{split}
\end{equation}
where $f_c$ is the carrier frequency, $c$ is the light speed, $\mu_{\mathsf{LoS}}$ and $\mu_{\mathsf{NLoS}}$ are the additional losses caused by the LoS and non-LoS link, respectively, which depend on the different environments, density, and height of buildings. $\widetilde{P}^{\mathsf{LoS}}_{ji,n}$ is a Bernoulli random variable that indicates whether the channel between the $i$-th sensor in the $j$-th cluster and the UAV is a line-of-sight (LoS) link, defined as $\widetilde{P}^{\mathsf{LoS}}_{ji,n} \sim \mathcal{B}(P^{\text{LoS}}_{ji,n})$, where $P^{\text{LoS}}_{ji,n}$ denotes the probability that the corresponding channel is a LoS channel, given by
\begin{equation}\label{eq:plos}
P^{\mathsf{LoS}}_{ji,n} = \frac{1}{1 + \varepsilon_1 \cdot\exp\left(-\varepsilon_2({\frac{180}{\pi}\arcsin\frac{H_n}{\|\bm{q}^\text{u}_n-\bm{q}_{ji}\|}}-\varepsilon_1)\right)},
\end{equation}
with $ \varepsilon_1 $ and $ \varepsilon_2 $ determined by the propagation environment. Accordingly, the channel power gain between the $i$-th sensor and the UAV in the $n$-th slot is given by
\begin{equation}\label{eq:gmn}
	\bm{g}_{ji,n} = 10^{-G_{ji,n} / 10}\tilde{\bm{g}}_{ji,n},~\forall j \in \mathcal{J},~\forall n \in \mathcal{N},
\end{equation}
where $\tilde{\bm{g}}_{ji,n}\in \mathbb{C}^{K\times 1}$ is the small-scale fading component.

Since AirComp is inherently sensitive to the propagation environment, we assume that the UAV communicates with only one cluster in a time slot to avoid inter-cluster interference. In particular, we define cluster scheduling as a set of binary variables $\left\{c_{j,n}\in \{0, 1\}| j \in \mathcal{J}, n \in \mathcal{N}\right\}$, where $c_{j,n}=1$ indicates that the $j$-th cluster is scheduled in the $n$-th slot, and $c_{j,n}=0$ otherwise. Due to differences in channel fading among intra-cluster sensors, signals from those in deep fading may fail to align with the expected amplitude at the receiver in a single time slot, causing computational distortion. To address this issue, we employ a multi-slot framework for intra-cluster AirComp. Then, we define the scheduling state of intra-cluster sensors as a set of binary variables $\left\{a_{ji,n}\in \{0, 1\}| j \in \mathcal{J}, i \in \mathcal{L}_j, n \in \mathcal{N}\right\}$, where $a_{ji,n}=1$ indicates that the $i$-th sensor in the $j$-th cluster is scheduled in the $n$-th slot, and $a_{ji,n}=0$ otherwise. The scheduling variables are imposed with the following constraints.
\begin{equation}
	\textstyle\sum\limits_{j \in \mathcal{J}}{c_{j,n} = 1},\quad\forall n \in \mathcal{N},
\end{equation}
\begin{equation}
	\textstyle\sum\limits_{n \in \mathcal{N}}{a_{ji,n} = 1},\quad\forall j\in\mathcal{J},~\forall i \in \mathcal{L}_j.
\end{equation}

\subsection{Multi-Slot AirComp Model}
For each sensor, the transmitted signal is derived from pre-processing of its raw sensing data. Let the sensing data symbol of the $i$-th sensor in the $j$-th cluster be $e_{ji} \in \mathbb{C}$, and the corresponding pre-processing signal is expressed as $o_{ji}=\Psi_j \left(e_{ji}\right)$, where $\Psi_j(\cdot)$ is the pre-processing function at the $j$-th cluster, used to transform the raw sensing data into normalized transmit symbols suitable for over-the-air aggregation. To facilitate the power-control design and reduce transmission power, the symbols from AirComp are assumed to be independent and normalized with zero mean and unit variance, i.e., $\mathbb{E}[o_{ji}] = 0$, $\mathbb{E}[o_{ji} \overline{o}_{ji}] = 1$. Meanwhile, the data among sensors in same cluster are uncorrelated, i.e., $\mathbb{E}[o_{ji} \overline{o}_{j\iota}]=0, \iota \neq i$, as in most of the existing studies~\cite{laws, uncor}. Based on the above signal model, the signal received at the UAV in the $n$-th slot can be modeled as
\begin{equation}
	y^{\text{R}}_{n} = \bm{w}^{\text{H}}_n\textstyle\sum\limits_{j \in \mathcal{J}}{c_{j,n}\textstyle\sum\limits_{i \in \mathcal{L}_j}a_{ji,n}b_{ji,n}\bm{g}_{ji,n} o_{ji}} + \bm{w}^{\text{H}}_n \bm{z}_n,~\forall n \in \mathcal{N},
\end{equation}
where $\bm{w}_n \in \mathbb{C}^{K\times1}$ with $\bm{w}^{\text{H}}_n \bm{w}_n=1$ is the aggregation beamforming vector, $(\cdot)^{\text{H}}$ stands for the conjugate transpose of the argument, $b_{ji,n} \in \mathbb{C}$ is the transmit coefficient, and $\bm{z}_n\in \mathbb{C}^{K\times1}$ is the additive white Gaussian noise vector with independent and identically distributed $\mathcal{CN}\left(0, \sigma^2\right)$. Furthermore, each sensor has a transmit power budget, and the peak transmission power constraint for the sensor is given by
\begin{equation}
	|b_{ji,n}|^2 \leq p_{\max},\quad\forall j\in\mathcal{J},~\forall i \in \mathcal{L}_j,~\forall n \in \mathcal{N},
\end{equation}
where $p_{\max}$ is the maximum transmit power, which is assumed identical across all sensors.

For data aggregation at the UAV, the sum of the distributed signals from all sensors corresponds to the superposition that naturally occurs in the channel, and the UAV aims at computing outputs for each cluster, called the target function is given by
\begin{equation}
	\hat{F}_j = \Phi_j \left(\textstyle\sum\limits_{i \in \mathcal{L}_j}o_{ji}\right),\quad \forall j\in\mathcal{J},
\end{equation}
where $\Phi(\cdot)$ is post-processing at the UAV. Without loss of generality, we consider the average of distributed data as the post-processing, which is expressed as $\hat{F}_j = \frac{1}{L_j}\left(\sum_{i \in \mathcal{L}_j}{o_{ji}}\right)$, but the proposed method can be extended to other nomographic functions~\cite{nomo}.

In practice, the signals received by the UAV during the $n$-th slot from the $j$-th cluster can be given as
\begin{equation} \label{eq:yr}
	Y_{j,n} = \bm{w}^{\text{H}}_n c_{j,n}\textstyle\sum\limits_{i \in \mathcal{L}_j}\left(a_{ji,n}b_{ji,n}\bm{g}_{ji,n} o_{ji} + \bm{z}_n\right).
\end{equation}
After $N$ slots, the sum of signals collected from the $j$-th cluster by the UAV can be given as
\begin{equation} \label{eq:Yr}
	Y_{j} = \textstyle\sum\limits_{n \in \mathcal{N}}\eta_n Y_{j,n},
\end{equation}
where $\eta_n \in \mathbb{R}_{+}$ is a receive normalizing factor at the UAV for signal power compensation and noise suppression at slot $n$. Furthermore, the data for each cluster is post-processed to obtain the corresponding estimated function, given by
\begin{equation}
	F_j = \frac{Y_{j}}{L_j}.
\end{equation}

We evaluate the AirComp error of the $j$-th cluster using the MSE between the estimated function $F_j$ and the target function $\hat{F}_j$. Specifically, considering the scheduling of clusters and sensors within clusters, the target function $\hat{F}_j$ can be represented as
\begin{equation}
	\hat{F}_j = \frac{1}{L_j} \left(\textstyle\sum\limits_{n \in \mathcal{N}}{c_{j,n}\textstyle\sum\limits_{i \in \mathcal{L}_j}a_{ji,n}o_{ji}}\right),\quad \forall j\in\mathcal{J}.
\end{equation}
Then, the MSE of the $j$-th cluster is given as
\begin{equation} \label{eq:md}
	\begin{split}
		\text{MSE}_j &= \mathbb{E}\left[|F_j- \hat{F}_j|^2\right] \\
		&= \frac{1}{L_j^2}\mathbb{E}\left[\left| Y_j- \textstyle\sum\limits_{n \in \mathcal{N}}c_{j,n}\textstyle\sum\limits_{i \in \mathcal{L}_j}a_{ji,n}{o_{ji}}\right| ^2\right],
	\end{split}
\end{equation}
where the expectation $\mathbb{E}[\cdot]$ is taken over the distributions of the transmitted signals and noise. Substituting~(\ref{eq:yr}) into~(\ref{eq:md}), the MSE can be further simplified as
\begin{equation}
	\begin{split}
		\text{MSE}_j =\:&\frac{1}{L_j^2}\mathbb{E}\left[\left| \textstyle\sum\limits_{n \in \mathcal{N}}c_{j,n}\textstyle\sum\limits_{i \in \mathcal{L}_j}a_{ji,n}{\left(\eta_n\bm{w}^{\text{H}}_n\bm{g}_{ji,n}b_{ji,n}-1\right) o_{ji}} \right.\right.\\
		&\left.\left. + \textstyle\sum\limits_{n \in \mathcal{N}}c_{j,n}\eta_n\bm{w}^{\text{H}}_n\bm{z}_n\right| ^2\right] \\
		=\:&\frac{1}{L_j^2}\left(\textstyle\sum\limits_{n \in \mathcal{N}}c_{j,n}\textstyle\sum\limits_{i \in \mathcal{L}_j}a_{ji,n}\left| \eta_n\bm{w}^{\text{H}}_n\bm{g}_{ji,n}b_{ji,n} - 1\right| ^2\right.\\
		&\left. +\textstyle\sum\limits_{n \in \mathcal{N}}c_{j,n}\eta_n^2\bm{w}^{\text{H}}_n \bm{w}_n\sigma^2\right).
	\end{split}
\end{equation}

\subsection{Energy Model}
Due to the fact that both the UAV and the sensors are energy-limited, we aim to achieve energy-efficient data computation in the considered system. Specifically, the UAV is powered by a limited on-board battery that must support flight, communication, and computation tasks, while the sensors are low-power devices with limited battery capacity. To improve the energy efficiency of the UAV-assisted AirComp system, both the propulsion energy and communication energy need to be addressed, as the former supports the UAV flying and the latter data transmissions. Therefore, we discuss the UAV propulsion energy consumption and the sensors communication energy consumption as follows.

For the AirComp transmissions of sensors, each sensor only transmits data in one time slot under the scheduling strategy. Therefore, the total transmission energy consumption of the AirComp system is given by
\begin{equation}
	E^{\text{tran}} = \textstyle\sum\limits_{n \in \mathcal{N}}\textstyle\sum\limits_{j \in \mathcal{J}}c_{j,n}\textstyle\sum\limits_{i \in \mathcal{L}_j}a_{ji,n}|b_{ji,n}|^2\delta.
\end{equation}

In consistence with the slotted aircomp operation, the UAV also constantly update its waypoints in each slot, the flight speed of the UAV between two successive waypoints can be regarded a constant and represented as
\begin{equation}
	v_n = \frac{\left\|\bm{q}^\text{u}_n-\bm{q}^\text{u}_{n-1}\right\|}{\delta},~\forall n\in\mathcal{N}.
\end{equation}
In addition, it must satisfy the maximum speed constraint:
\begin{equation}
	v_n \leq v_{\max},~\forall n\in\mathcal{N}.
\end{equation}
The UAV flight energy is dominated by its attributes and flight speed. Therefore, the flight power consumption of the rotary-wing UAV can be expressed as
\begin{equation}
	\begin{split}
	P(v_n) =\:& P_0\left(1+\frac{3v_n^2}{W^2_{\text{tip}}}\right) + P_i\left(\sqrt{1+\frac{v_n^4}{4v^4_0}}-\frac{v_n^2}{2v^2_0}\right)^{1/2}\\
	&+ \frac{1}{2}d_0\rho \omega\Lambda v_n^3,~\forall n\in\mathcal{N},
	\end{split}	
\end{equation}
where $P_0$ and $P_i$ are constants representing the blade profile power and induced power while hovering, $W_{\text{tip}}$ is the tip speed of the rotor blade, $v_0$ is the mean rotor induced velocity in hover, $d_0$ and $\omega$ are the fuselage drag ratio and rotor solidity, respectively, and $\rho$ and $\Lambda$ denote the air density and rotor discarea, respectively. Here, we ignore the energy consumption of the UAV during vertical ascent, descent, and acceleration/deceleration to simplify the problem analysis. Consider the whole flying journey, the propulsion energy consumption can be given as
\begin{equation}
	E^{\text{prop}} = \textstyle\sum\limits_{n \in \mathcal{N}}P(v_n)\delta.
\end{equation}

\subsection{Problem Formulation} \label{sec:prob}
In the considered UAV-assisted multi-cluster AirComp network, mitigating AirComp errors while reducing system energy consumption stands as a paramount objective. Notably, signal misalignment arising from heterogeneous channel conditions among intra-cluster sensors can severely compromise AirComp accuracy, which motivates the design of AirComp transceiving and scheduling strategy to compensate for fading effects. Moreover, UAV trajectory exerts a direct influence on both channel quality and flight energy expenditure, creating a pronounced interplay between computational precision and system sustainability. Accordingly, the system design challenge naturally evolves into a joint optimization of AirComp transceiving strategy, scheduling strategy, and trajectory planning, aiming to reduce computational errors and system energy consumption under the constraints of the multi-slot AirComp framework. Mathematically, the joint optimization problem is formulated as follows:
\begin{IEEEeqnarray}{cl}
	\IEEEyesnumber\label{eq:p0} \IEEEyessubnumber*
	\min_{\begin{subarray}{c}
			\bm{B}, \bm{C}, \bm{A}, \\
			\left\{\eta_n, \bm{w}_n, \right.\\
			\left.\bm{q}^{\text{u}}_n\right\}_{n\in \mathcal{N}}
	\end{subarray}} &~\kappa_1\textstyle\sum\limits_{j\in \mathcal{J}}\text{MSE}_j +\kappa_2 \left(E^{\text{tran}}+\varsigma E^{\text{prop}}\right)   \label{eq:ob0} \\ 
	\rm{s.t.} & |b_{ji,n}|^2 \leq p_{\max},\:\forall j\in\mathcal{J}, \forall i\in\mathcal{L}_j, \forall n\in\mathcal{N},\quad\label{eq:pmax}\\
	& \eta_n > 0,~\forall n \in \mathcal{N},\label{eq:eta} \\
	& \bm{w}^{\text{H}}_n \bm{w}_n=1,~\forall n \in \mathcal{N}, \label{eq:vv} \\
	&\textstyle\sum\limits_{j \in \mathcal{J}}{c_{j,n} = 1},~\forall n \in \mathcal{N},  \label{eq:c0}\\
	&\textstyle\sum\limits_{n \in \mathcal{N}}{a_{ji,n} = 1},~\forall j\in\mathcal{J}, i\in\mathcal{L}_j,  \label{eq:a0}\\
	& c_{j,n}, a_{ji,n} \in \{0, 1\},\forall j\in\mathcal{J}, \forall i\in\mathcal{L}_j, \forall n\in\mathcal{N},\quad\label{eq:ca}\\
	& \bm{q}^\text{u}_0 = \bm{q}_o,~\bm{q}^\text{u}_{N} = \bm{q}_f,\\
	& v_n \leq v_{\max},~\forall n\in\mathcal{N},\\
	& H_{\min} \leq H_n \leq H_{\max},~\forall n\in\mathcal{N},
\end{IEEEeqnarray}
where $\bm{B}=\left\{\bm{B}_n\right\}_{n\in\mathcal{N}}$, $\bm{B}_n=\left\{b_{ji,n}\right\}_{j\in\mathcal{J}, i\in\mathcal{L}_j}$, $\bm{C}=\left\{c_{j,n}\right\}_{j\in\mathcal{J}, n\in\mathcal{N}}$, and $\bm{A}=\left\{a_{ji,n}\right\}_{j\in\mathcal{J}, i\in\mathcal{L}_j, n\in\mathcal{N}}$. $\kappa_1$ and $\kappa_2$ are the trade-off coefficients for AirComp accuracy and energy consumption, respectively, with $\kappa_1 + \kappa_2 = 1$. $\varsigma$ is the weighting factor of energy.

For the formulated problem, the considered factors affect system performance in a coupled and complicated manner. Specifically, due to the integer scheduling variables $a_{ji, n}$ and $c_{j,n}$, the problem in~(\ref{eq:p0}) involves an integer objective function and integer constraints in~(\ref{eq:c0}),~(\ref{eq:a0}), and~(\ref{eq:ca}). Meanwhile, since the scheduling strategy directly affects the AirComp transceiving strategy, the continuous variables $b_{ji,n}$, $\eta_{n}$, $\bm{w}_{n}$ and integer scheduling variables become coupled in both computation error and transmission energy consumption, thereby leading to the non-convexity of the objective function. In addition, the UAV trajectory design further interacts with both the scheduling and transmission strategies, since the UAV's position determines the channel conditions, which in turn influence the optimal scheduling and transmission decisions. Thus, the problem in~(\ref{eq:p0}) is a non-convex problem with mixed-integer nonlinear programming, which is cumbersome to reach the optimum.

Revisit the formulated problem, we can observe that the transmission coefficients of sensors, the aggregation beamforming vector, and the normalizing factor can be investigated in a slot-wise manner with a fixed scheduling strategy. In contrast, the scheduling strategy and the UAV trajectory affect the system performance across all slots, and the strategies between each slot influence each other. Therefore, this problem can be decomposed into two layers, where the outer layer solves the scheduling strategy and the UAV trajectory planning, while the inner layer is for the AirComp transceiving strategy optimization. The inner problem can be solved independently in each slot, while the outer problem is solved through reinforcement learning processing for scheduling strategy and the UAV trajectory.

\section{Alternating Optimization for AirComp Transceiving Strategy} \label{sec:inner}
In this section, we consider the inner problem for AirComp transceiving strategy optimization, with a fixed scheduling strategy and UAV trajectory at the outer layer. The inner problem can be investigated in a slot-wise manner, aiming to maximize the energy efficiency of AirComp in a single time slot. This enables a structured signal transmission module that can be repeatedly called in the outer scheduling-trajectory optimization loop. In particular, the inner problem at slot $n$ can be described as follows
\begin{IEEEeqnarray}{cl}
	\IEEEyesnumber\label{eq:p1} \nonumber
	\min_{\bm{B}_n, \eta_n, \bm{w}_n} & \:\kappa_1\textstyle\sum\limits_{j \in \mathcal{J}}\frac{c_{j,n}}{L_j^2}\bigg(\textstyle\sum\limits_{i \in \mathcal{L}_j}a_{ji,n}\left|\eta_n\bm{w}^{\text{H}}_n\bm{g}_{ji,n}b_{ji,n}- 1\right|^2\:\bigg.\quad\\
	\bigg.&+ \eta_n^2\bm{w}^{\text{H}}_n \bm{w}_n\sigma^2\bigg) + \kappa_2 \textstyle\sum\limits_{j \in \mathcal{J}}c_{j,n}\textstyle\sum\limits_{i \in \mathcal{L}_j}a_{ji,n}|b_{ji,n}|^2\delta\quad\:\\ \nonumber
	\rm{s.t.} \: &~(\text{\ref{eq:pmax}}),~(\text{\ref{eq:eta}}),~(\text{\ref{eq:vv}}).
\end{IEEEeqnarray}
However, the problem in~(\ref{eq:p1}) remains challenging to solve due to the non-concave objective function involving the coupled variables. Therefore, we propose an alternating optimization framework to solve the problem in~(\ref{eq:p1}), by iteratively optimizing one of the transmit coefficients, the normalizing factor, and the aggregation beamforming vector with the other being fixed at each iteration as detailed below.

For given the normalizing factor $\eta_n$ and the aggregation beamforming vector $\bm{w}_n$, the problem in~(\ref{eq:p1}) can be transformed to the following form
\begin{IEEEeqnarray}{cl}
	\IEEEyesnumber\label{eq:p1-b} \nonumber
	\min_{\bm{B}_n} \quad & \kappa_1\textstyle\sum\limits_{j \in \mathcal{J}}\frac{c_{j,n}}{L_j^2}\textstyle\sum\limits_{i \in \mathcal{L}_j}a_{ji,n}\left|\eta_n\bm{w}^{\text{H}}_n\bm{g}_{ji,n}b_{ji,n} - 1\right| ^2\\
	&+\kappa_2 \textstyle\sum\limits_{j \in \mathcal{J}}c_{j,n}\textstyle\sum\limits_{i \in \mathcal{L}_j}a_{ji,n}|b_{ji,n}|^2\delta  \\  \nonumber
	\rm{s.t.} \quad &(\text{\ref{eq:pmax}}).
\end{IEEEeqnarray}
Meanwhile, when the transmit coefficients $\bm{B}_n$ and the aggregation beamforming vector $\bm{w}_n$ are given, we optimize the normalizing factor $\eta_n$. Accordingly, the problem in~(\ref{eq:p1}) is reduced to the following form
\begin{IEEEeqnarray}{cl}
	\IEEEyesnumber\label{eq:p1-e}
	\min_{\begin{subarray}{c}
			\eta_n \nonumber
	\end{subarray}} \quad & \kappa_1\textstyle\sum\limits_{j \in \mathcal{J}}\frac{c_{j,n}}{L_j^2}\bigg(\textstyle\sum\limits_{i \in \mathcal{L}_j}a_{ji,n}\left| \eta_n\bm{w}^{\text{H}}_n\bm{g}_{ji,n}b_{ji,n}- 1\right| ^2\bigg.\\
	&\bigg.+\eta_n^2\sigma^2\bigg)\\ \nonumber
	\rm{s.t.} \quad &~(\text{\ref{eq:eta}}).
\end{IEEEeqnarray}
Finally, for the given transmit coefficients $\bm{B}_n$ and the normalizing factor $\eta_n$, we optimize the aggregation beamforming vector $\bm{w}_n$. Then, the problem in~(\ref{eq:p1}) is recast as the following problem
\begin{IEEEeqnarray}{cl}
	\IEEEyesnumber\label{eq:p1-w}
	\min_{\begin{subarray}{c}
			\bm{w}_n \nonumber
	\end{subarray}} \: & \kappa_1\textstyle\sum\limits_{j \in \mathcal{J}}\frac{c_{j,n}}{L_j^2}\bigg(\textstyle\sum\limits_{i \in \mathcal{L}_j}a_{ji,n}\left|\eta_n\bm{w}^{\text{H}}_n\bm{g}_{ji,n}b_{ji,n}- 1\right|^2\:\bigg.\\
\bigg.&+ \eta_n^2\bm{w}^{\text{H}}_n \bm{w}_n\sigma^2\bigg)\\ \nonumber
	\rm{s.t.} \: &~(\text{\ref{eq:vv}}).
\end{IEEEeqnarray}

We then discuss the three individual subproblems of the aforementioned inner-layer problem as follows.

\textit{1) Optimization of the Transmit Coefficients:}
The problem in~(\ref{eq:p1-b}) is a quadratically constrained quadratic program. In this problem, the transmit coefficients to be optimized are mutually independent, which allows the problem to be decomposed into multiple subproblems, each involving only a single variable $b_{ji,n}$. For a sensor transmitting in the $n$-th slot (i.e., $c_{j,n}=1$ and $a_{ji,n}=1$), its transmit coefficient optimization problem can be written as follows:
\begin{IEEEeqnarray}{cl}
	\IEEEyesnumber\label{eq:p1-b1}
	\min_{\begin{subarray}{c}
			b_{ji,n}
	\end{subarray}} \quad & \frac{\kappa_1}{L_j^2}\left| \eta_n\bm{w}^{\text{H}}_n\bm{g}_{ji,n}b_{ji,n} - 1\right|^2+\kappa_2 |b_{ji,n}|^2\delta  \\ \nonumber
	\rm{s.t.} \quad &~(\text{\ref{eq:pmax}}).
\end{IEEEeqnarray}
It can be observed that the first term of objective function in~(\ref{eq:p1-b1}) follows that
\begin{equation}
	\begin{split}
		&\frac{\kappa_1}{L_j^2}\left| \eta_n\bm{w}^{\text{H}}_n\bm{g}_{ji,n}b_{ji,n} - 1\right|^2\\ &=\frac{\kappa_1}{L_j^2}\left[\left| \eta_n\bm{w}^{\text{H}}_n\bm{g}_{ji,n}b_{ji,n}\right|^2 - 2\Re\left( \eta_n\bm{w}^{\text{H}}_n\bm{g}_{ji,n}b_{ji,n}\right) + 1\right]\\
		&\geq \frac{\kappa_1}{L_j^2}\left[\left| \eta_n\bm{w}^{\text{H}}_n\bm{g}_{ji,n}\right|^2\left|b_{ji,n}\right|^2 - 2\left| \eta_n\bm{w}^{\text{H}}_n\bm{g}_{ji,n}\right|\left|b_{ji,n}\right| + 1\right],\quad
	\end{split}	
\end{equation}
where the equality holds if and only if $\eta_n\bm{w}^{\text{H}}_n\bm{g}_{ji,n}b_{ji,n}$ is real and non-negative. Therefore, to minimize the objective function in~(\ref{eq:p1-b1}), it is required that each term $\eta_n\bm{w}^{\text{H}}_n\bm{g}_{ji,n}b_{ji,n}$ be real and non-negative. Without loss of generality, we set $b_{ji,n}=\frac{\sqrt{p_{ji,n}}\left(\bm{w}^{\text{H}}_n\bm{g}_{ji,n}\right)^{\ast}}{\left|\bm{w}^{\text{H}}_n\bm{g}_{ji,n}\right|}, p_{ji,n}\in\left[0,p_{\max}\right]$ to cancel the phase shifts introduced by the complex channel coefficients, where $p_{ji,n}$ denotes the transmit power of the $i$-th sensor in the $j$-th cluster during slot $n$. Hence, the problem in~(\ref{eq:p1-b1}) can be reformulated as
\begin{IEEEeqnarray}{cl}
	\IEEEyesnumber\label{eq:p1-b2}
	\min_{\begin{subarray}{c}\IEEEyessubnumber*
			p_{ji,n}
	\end{subarray}} \quad & \frac{\kappa_1}{L_j^2}\left(\eta_n\left|\bm{w}^{\text{H}}_n\bm{g}_{ji,n}\right|\sqrt{p_{ji,n}} - 1\right) ^2+\kappa_2 p_{ji,n}\delta  \\
	\rm{s.t.} \quad &0\leq p_{ji,n} \leq p_{\max}, ~\forall j\in\mathcal{J}, \forall i\in\mathcal{L}_j.\label{eq:ppmax}
\end{IEEEeqnarray}
We can see that the problem in~(\ref{eq:p1-b2}) is a convex linearly constrained quadratic program. In this work, we solve this problem via the Lagrange duality method to obtain a more efficient solution. Let $\lambda_p \geq 0$ denote the dual variable associated with the constraint in~(\ref{eq:ppmax}). Then, the Lagrangian of the problem in~(\ref{eq:p1-b2}) is
\begin{equation}\label{eq:p1-b3}
	\begin{split}
		\mathcal{L}\left(p_{ji,n},\lambda_p\right) =\:&\frac{\kappa_1}{L_j^2}\left(\eta_n\left|\bm{w}^{\text{H}}_n\bm{g}_{ji,n}\right|\sqrt{p_{ji,n}} - 1\right) ^2 \\
		&+\kappa_2 p_{ji,n}\delta + \lambda_p\left(p_{ji,n}-p_{\max}\right).
	\end{split}	
\end{equation}
By applying the Karush–Kuhn–Tucker (KKT) conditions, we obtain the solution to the problem in~(\ref{eq:p1-b2}). Specifically, the solution can be derived by setting the first derivative of $\mathcal{L}\left(p_{ji,n},\lambda_p\right)$ with respect to $p_{ji,n}$ to zero, yielding
\begin{equation}\label{eq:p1-b4}
	p_{ji,n} =\left(\frac{\kappa_1\eta_n\left|\bm{w}^{\text{H}}_n\bm{g}_{ji,n}\right|}{\kappa_1\eta_n^2\left|\bm{w}^{\text{H}}_n\bm{g}_{ji,n}\right|^2 + \left(\kappa_2\delta + \lambda_p\right)L_j^2}\right)^2.
\end{equation}

If $\lambda_p > 0$, the maximum transmit power constraint of the $i$-th sensor in the $j$-th cluster in slot $n$ must be active at the optimum by the complementary slackness condition, i.e.,
\begin{equation}\label{eq:kkt}
	\lambda_p\left(p_{ji,n}-p_{\max}\right)=0.
\end{equation}
Therefore, combining~(\ref{eq:p1-b4}) and~(\ref{eq:kkt}), we obtain $p^{\star}_{ji,n}=p_{\max}$.

\begin{algorithm}
	\footnotesize   
	\caption{Alternating optimization for solving~(\ref{eq:p1})}
	\label{alg:1}
	\SetKwData{In}{\textbf{in}}\SetKwData{To}{to}
	\DontPrintSemicolon
	\SetAlgoLined
	\KwIn {$p_{\max}, \left\{c_{j,n}\right\}_{\forall j\in\mathcal{J}}, \left\{a_{ji,n}\right\}_{\forall j\in\mathcal{J}, \forall i\in\mathcal{L}_j}, \bm{q}^{\text{u}}_n$.}
	\KwOut {$\bm{B}_n$, $\eta_n$, $\bm{w}_n$.}
	{Initialization:~$l \leftarrow 0$, set $\eta^{(0)}_{n}$, $\bm{w}^{(0)}_n$}.\\
	\Repeat{\footnotesize$\|\bm{B}^{(l)}_n-\bm{B}^{(l-1)}_n\|_F+\|\eta^{(l)}_n-\eta^{(l-1)}_n\|+\|\bm{w}^{(l)}_n-\bm{w}^{(l-1)}_n \|\leq\Gamma$.}{$l \leftarrow l+1$;\\
		Given $\eta^{(l-1)}_n$ and $\bm{w}^{(l-1)}_n$, update $\bm{B}^{(l)}_n$ based to~(\ref{eq:p1-b8}).\\
		Given $\bm{B}^{(l)}_n$ and $\bm{w}^{(l-1)}_n$, update $\eta^{(l)}_n$ according to~(\ref{eq:p1-e1}).\\
		Given $\bm{B}^{(l)}_n$ and $\eta^{(l)}_n$, solve~(\ref{eq:p1-w9}) by the bisection method to obtain $\lambda_w$, then update $\bm{w}^{(l)}_n$ following to~(\ref{eq:p1-w4}).
	}
\end{algorithm}

If $\lambda_p = 0$, substituting $\lambda_p = 0$ into~(\ref{eq:p1-b4}), we can obtain that
\begin{equation}\label{eq:p1-b6}
	p_{ji,n} =\left(\frac{\kappa_1\eta_n\left|\bm{w}^{\text{H}}_n\bm{g}_{ji,n}\right|}{\kappa_1\eta_n^2\left|\bm{w}^{\text{H}}_n\bm{g}_{ji,n}\right|^2 + \kappa_2\delta L_j^2}\right)^2.
\end{equation}
Moreover, by the primal feasibility condition, the maximum transmit power constraint must satisfy $p_{ji,n}\leq p_{\max}$. Hence, the optimal $p_{ji,n}$ is given by
\begin{equation}\label{eq:p1-b7}
	p^{\star}_{ji,n} = \min\left\{\left(\frac{\kappa_1\eta_n\left|\bm{w}^{\text{H}}_n\bm{g}_{ji,n}\right|}{\kappa_1\eta_n^2\left|\bm{w}^{\text{H}}_n\bm{g}_{ji,n}\right|^2 + \kappa_2\delta L_j^2}\right)^2,\: p_{\max}\right\},
\end{equation}
and the corresponding optimal $b_{ji,n}$ is given by
\begin{equation}\label{eq:p1-b8}
	b^{\star}_{ji,n}=\frac{\sqrt{p^{\star}_{ji,n}}\left(\bm{w}^{\text{H}}_n\bm{g}_{ji,n}\right)^{\ast}}{\left|\bm{w}^{\text{H}}_n\bm{g}_{ji,n}\right|},
\end{equation}
where $(\cdot)^{\ast}$ stands for the conjugate of the argument.

\textit{2) Optimization of the Normalizing Factor:}
The problem in~(\ref{eq:p1-e}) is a quadratic program. By setting the first derivative of the objective function in~(\ref{eq:p1-e}) to zero, its closed-form solution can be obtained as follows.

\begin{equation}\label{eq:p1-e1}
	\eta^{\star}_n = \textstyle\sum\limits_{j \in \mathcal{J}}c_{j,n}\frac{\textstyle\sum\limits_{i \in \mathcal{L}_j}a_{ji,n}\left|\bm{w}^{\text{H}}_n\bm{g}_{ji,n}b_{ji,n}\right|}{\textstyle\sum\limits_{i \in \mathcal{L}_j}a_{ji,n}\left| \bm{w}^{\text{H}}_n\bm{g}_{ji,n}b_{ji,n}\right| ^2 + \sigma^2}.
\end{equation}

\textit{3) Optimization of the Aggregation Beamforming:}
The problem in~(\ref{eq:p1-w}) is still challenging due to the non-convex norm constraint on $\bm{w}_n$. For ease of analysis, we consider the case where the sensors in the $j$-th cluster transmit during the $n$-th slot (i.e., $c_j=1$). Accordingly, the problem in~(\ref{eq:p1-w}) can be reformulated as
\begin{IEEEeqnarray}{cl}
	\IEEEyesnumber\label{eq:p1-w1}
	\min_{\begin{subarray}{c}
			\bm{w}_n
	\end{subarray}} \quad & \textstyle\sum\limits_{i \in \mathcal{L}_j}a_{ji,n}\left| \eta_n\bm{w}^{\text{H}}_n\bm{g}_{ji,n}b_{ji,n}- 1\right| ^2\\  \nonumber
		\rm{s.t.} \quad &~(\text{\ref{eq:vv}}).
\end{IEEEeqnarray}
Let $\lambda_w$ denote the dual variable associated with the constraint in~(\ref{eq:vv}). Then, the Lagrangian of the problem in~(\ref{eq:p1-w1}) is
\begin{equation}\label{eq:p1-w2}
	\begin{split}
		\mathcal{L}\left(\bm{w}_n,\lambda_w\right) = &\textstyle\sum\limits_{i \in \mathcal{L}_j}a_{ji,n}\left| \eta_n\bm{w}^{\text{H}}_n\bm{g}_{ji,n}b_{ji,n}- 1\right| ^2 \\
		&+ \lambda_w\left(\bm{w}^{\text{H}}_n \bm{w}_n-1\right).
	\end{split}	
\end{equation}
Then, by setting the first derivative of $\mathcal{L}\left(\bm{w}_n,\lambda_w\right)$ with respect to $\bm{w}_n$ to zero
\begin{equation}\label{eq:p1-w3}
	\begin{split}
		\nabla\mathcal{L}\left(\bm{w}_n,\lambda_w\right) = &\left(\textstyle\sum\limits_{i \in \mathcal{L}_j}a_{ji,n} \left|\eta_n b_{ji,n}\right|^2\bm{g}_{ji,n}\bm{g}^{\text{H}}_{ji,n}+\lambda_w\bm{\mathbf{I}}\right)\bm{w}_n \\
		&- \textstyle\sum\limits_{i \in \mathcal{L}_j}a_{ji,n}\eta_n b_{ji,n}\bm{g}_{ji,n}=0,
	\end{split}	
\end{equation}
we obtain
\begin{equation}\label{eq:p1-w4}
	\begin{split}
	\bm{w}_n \left(\lambda_w\right) = & \Big(\textstyle\sum\limits_{i \in \mathcal{L}_j}a_{ji,n}\left|\eta_nb_{ji,n}\right|^2\bm{g}_{ji,n}\bm{g}^{\text{H}}_{ji,n}+\lambda_w\bm{\mathbf{I}}\Big)^{-1} \\
	&\textstyle\sum\limits_{i \in \mathcal{L}_j}a_{ji,n}\eta_nb_{ji,n}\bm{g}_{ji,n}.
	\end{split}
\end{equation}
To simplify the notation, we define the two constant terms in the $\bm{w}_n \left(\lambda_w\right)$ function as
\begin{equation}\label{eq:p1-w5}
	\bm{\Theta}_n = \textstyle\sum\limits_{i \in \mathcal{L}_j}a_{ji,n}\left|\eta_nb_{ji,n}\right|^2\bm{g}_{ji,n}\bm{g}^{\text{H}}_{ji,n},
\end{equation}
and
\begin{equation}\label{eq:p1-w6}
	\bm{\Omega}_n = \textstyle\sum\limits_{i \in \mathcal{L}_j}a_{ji,n}\eta_nb_{ji,n}\bm{g}_{ji,n}.
\end{equation}
Furthermore, we need to find a $\lambda_w$ such that $\bm{w}_n \left(\lambda_w\right)$ satisfies the constraint in~(\ref{eq:vv}). Substituting the equation in~(\ref{eq:p1-w4}) into the constraint~(\ref{eq:vv}), we obtain
\begin{equation}\label{eq:p1-w7}
		\bm{\Omega}^{\text{H}}_n\Big(\bm{\Theta}_n+\lambda_w\bm{\mathbf{I}}\Big)^{-2}\bm{\Omega}_n=1.
\end{equation}
Since $\bm{\Theta}_n$ is an Hermitian matrix, it can be eigen-decomposed as
\begin{equation}\label{eq:p1-w8}
	\bm{\Theta}_n=\bm{U}_n\bm{\Xi}_n\bm{U}^{\text{H}}_n,
\end{equation}
where $\bm{U}_n$ is a unitary matrix, $\bm{\Xi}_n= \mathbf{diag}\left(\tau_1, \tau_2, \cdots, \tau_K\right)$ is an eigenvalue matrix, and the eigenvalue $\tau_k \geq 0$. Then, the equation in~(\ref{eq:p1-w7}) can be rewritten as
\begin{equation}\label{eq:p1-w9}
	\textstyle\sum\limits_{k \in K}\frac{\left|\bm{U}^{\text{H}}_n\bm{\Omega}_n\right|^2}{\left(\tau_k+\lambda_w\right)^2}=1,
\end{equation}
which can be solved using root-finding methods (e.g., the Newton-Raphson method or the bisection method). Substituting the optimal solution $\lambda^{\star}_w$ into~(\ref{eq:p1-w4}) yields the optimal aggregation beamforming vector $\bm{w}_n$.

\textit{4) Alternating Optimization Algorithm:}
Based on the decomposition of the aforementioned inner-layer problem, the original optimization problem in~(\ref{eq:p1}) can be tackled through an alternating optimization algorithm involving~(\ref{eq:p1-b8}),~(\ref{eq:p1-e1}), and~(\ref{eq:p1-w4}), which is summarized in Alg.~\ref{alg:1}. Specifically, with fixed $\eta_n$ and $\bm{w}_n$, the optimized transmit coefficients $\bm{B}^{\star}_n$ are obtained from~(\ref{eq:p1-b8}). Then, given the fixed $\bm{w}_n$ and the optimized transmit coefficients $\bm{B}^{\star}_n$, an optimized normalizing factor $\eta^{\star}_n$ is obtained from~(\ref{eq:p1-e1}). Next, using the optimized $\bm{B}^{\star}_n$ and $\eta^{\star}_n$, the optimized aggregation beamforming $\bm{w}^{\star}_n$ is obtained from~(\ref{eq:p1-w4}) and~(\ref{eq:p1-w9}). Finally, by iteratively alternating until convergence accuracy $\Gamma$ is met, the solution to the original problem is achieved.

\section{Pointer Network-Assisted SAC Learning for Scheduling and UAV Trajectory} \label{sec:outer}
In this section, we consider the outer problem of scheduling and UAV trajectory, with the aircomp transceving strategy provided through inner-layer optimization. Unlike the inner problem, which can be solved at each slot, the UAV trajectory and scheduling impact sensor transmission and AirComp throughout the entire mission time. The mutual influence of UAV trajectory, scheduling strategy, and inner-layer AirComp strategy leads to significant complexity in the problem-solving. This complexity renders traditional optimization algorithms difficult in the solution-seeking process. In this regard, DRL emerges as an attractive solution~\cite{sgdrl}.

However, for our considered problem with a high dimensional state space and the coexistence of discrete scheduling decisions and continuous trajectory control, the traditional DRL may face the following issues~\cite{dedrl}. On one hand, scheduling is a combinatorial problem that requires sequential cluster and sensor selection, and traditional continuous-control algorithms such as deep deterministic policy gradient and SAC cannot effectively capture its permutation structure. On the other hand, trajectory optimization relies on continuous dynamics, where discrete-action methods such as deep Q-network and proximal policy optimization with discretization suffer from poor scalability. Moreover, on-policy methods such as Proximal Policy Optimization generally require a large amount of freshly collected interaction data for policy updates, which makes them less suitable for our setting, where each environment interaction is additionally coupled with inner-layer transceiver optimization. As a result, conventional single-layer DRL methods are insufficient for our problem. To address these challenges, we propose a pointer network-assisted DRL framework, where the cluster and sensor scheduling is optimized using an actor-critic architecture with a pointer network as the actor, while the SAC algorithm carries out UAV trajectory optimization. Within each SAC training episode, the pointer network first generates a scheduling strategy, and then the trajectory is optimized accordingly, with the pointer network updated periodically. This hierarchical design enables coordinated improvements between scheduling and trajectory optimization, thereby improving availability compared with conventional DRL approaches.

\begin{figure*}[t]
        \centering
        \includegraphics[width=0.98\linewidth]{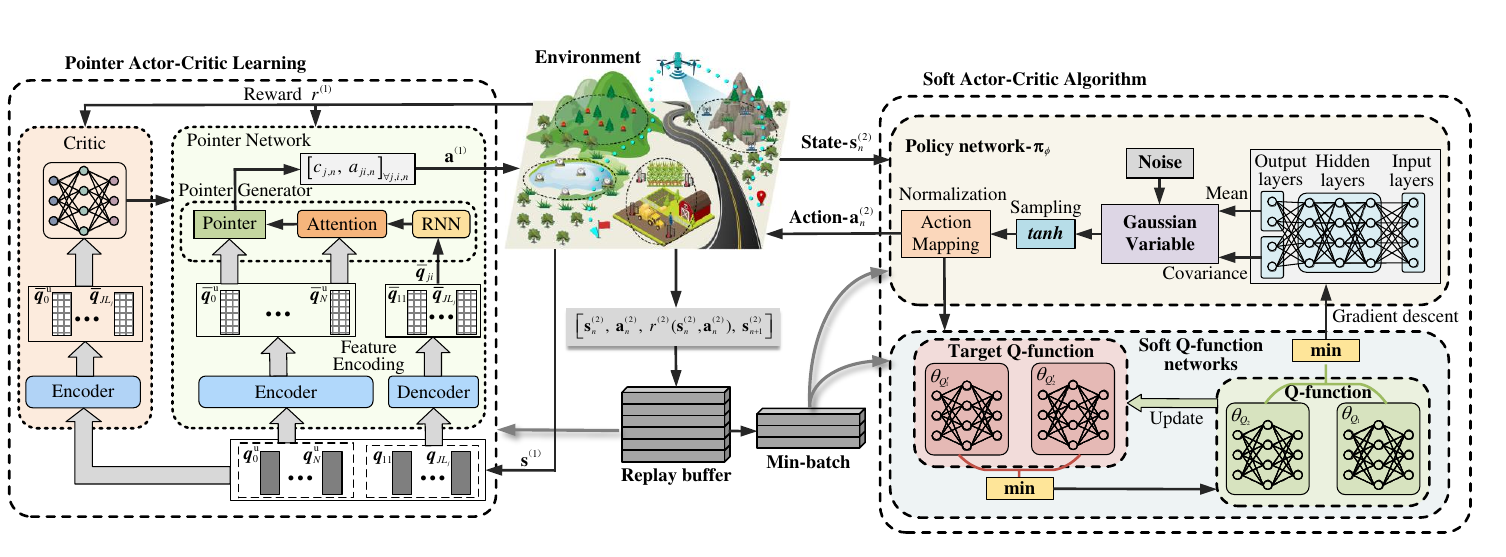}
		\vspace{-3.5mm}
        \caption{The framework of pointer network-assisted SAC learning algorithm.}
        \vspace{-3.5mm}
        \label{fig:pn-sac}
\end{figure*}

\subsection{Trajectory and Scheduling as MDP} \label{sec:MDP}
The UAV trajectory planning and the scheduling strategy have long-term implications on AirComp and energy consumption during missions. To address these long-term effects, the UAV can be treated as an agent, with trajectory planning and scheduling strategy as the action space, and AirComp and energy consumption states as the state space, modeled as a Markov decision process (MDP) to track the long-term effects. From a learning perspective, the problem can be defined as policy search in two interacting MDPs, for sensor scheduling and for UAV trajectory control, respectively. Specifically, the scheduling MDP generates scheduling decisions based on the latest UAV trajectory provided by the trajectory MDP, while the trajectory MDP provides the UAV trajectory according to the scheduling strategy determined by the scheduling MDP. Let the tuple $\left(\mathcal{S}^{\scriptscriptstyle(1)}, \mathcal{A}^{\scriptscriptstyle(1)}, r^{\scriptscriptstyle(1)}\right)$ denote the scheduling MDP and a tuple $\left(\mathcal{S}^{\scriptscriptstyle(2)}, \mathcal{A}^{\scriptscriptstyle(2)}, r^{\scriptscriptstyle(2)}\right)$ denote the trajectory MDP. $\mathcal{S}^{\scriptscriptstyle(1)}$ and $\mathcal{S}^{\scriptscriptstyle(2)}$ denote the state spaces of the scheduling and trajectory MDPs, $\mathcal{A}^{\scriptscriptstyle(1)}$ and $\mathcal{A}^{\scriptscriptstyle(2)}$ denote their corresponding action spaces, and $r^{\scriptscriptstyle(1)}$ and $r^{\scriptscriptstyle(2)}$ are the respective reward functions. Specifically, the corresponding elements are defined as follows.

\begin{algorithm}
	\caption{Pointer Network-Assisted SAC Learning}
	\label{alg:2}
	\SetKwData{In}{\textbf{in}}\SetKwData{To}{to}
	\DontPrintSemicolon
	\SetAlgoLined
	\KwIn {$p_{\max}, v_{\max}, H_{\min}, H_{\max}$, $\bm{q}_0$, and $\bm{q}_f$.}
	\KwOut {$\bm{B}, \bm{C}, \bm{A},	\left\{\eta_n, \bm{w}_n, \bm{q}^{\text{u}}_n\right\}_{n\in \mathcal{N}}$.}
	{Initialization: The parameters of policy networks, Q-function networks, and target Q-function networks: $\phi_{\text{p}}$, $\phi$, $\theta_{\text{p}}$, $\theta_{1}$, and $\theta_{2}$, $\theta'_{1} \gets \theta_{1}$, $\theta'_{2} \gets \theta_{2}$; The initial scheduling strategy $\left\{\widetilde{\bm{C}},\widetilde{\bm{A}}\right\}$.}\\
	\For{\rm{each episode}}{
		Obtain initial state $\mathbf{s}_1^{\scriptscriptstyle(2)}$ based on $\left\{\widetilde{\bm{C}},\widetilde{\bm{A}}\right\}$ generated by $\pi_{\phi_{\text{p}}}$ and $\left\{\bm{B}_1,\eta_1,\bm{w}_1\right\}$ optimized by Alg.~\ref{alg:1}; \;
		\For{\rm{slot} $n=1,2,\cdots,N$}{
			Sample action from policy $\mathbf{a}_{n}^{\scriptscriptstyle(2)} \thicksim\pi_{\phi}(\mathbf{a}_n^{\scriptscriptstyle(2)}|\mathbf{s}_n^{\scriptscriptstyle(2)})$; \;
			Execute action $\mathbf{a}_{n}^{\scriptscriptstyle(2)}$, optimize $\left\{\bm{B}_n,\eta_n,\bm{w}_n\right\}$ based on Alg.~\ref{alg:1}, observe next state $\mathbf{s}_{n+1}^{\scriptscriptstyle(2)}$, and receive reward $r^{\scriptscriptstyle(2)}(\mathbf{s}_{n}^{\scriptscriptstyle(2)}, \mathbf{a}_{n}^{\scriptscriptstyle(2)})$; \;
			\If{n = N \rm{and} $\sqrt{\left\|\bm{q}^{\text{u}}_n-\bm{q}_f\right\|^2} \leq v_{\max}\delta$}{
				Achieve reward $r_f$; \;
			}
			Store $(\mathbf{s}_{n}^{\scriptscriptstyle(2)}, \mathbf{a}_{n}^{\scriptscriptstyle(2)}, r^{\scriptscriptstyle(2)}(\mathbf{s}_{n}^{\scriptscriptstyle(2)}, \mathbf{a}_{n}^{\scriptscriptstyle(2)}),\mathbf{s}_{n+1}^{\scriptscriptstyle(2)})$ into the replay buffer $\mathcal{D}$;
		}	
		\If{\rm{episode $\mathsf{mod}$ training interval = 0}}{Retrieve new UAV trajectories (one training interval) from $\mathcal{D}$; \;
			Generate action $\mathbf{a}^{\scriptscriptstyle(1)}$ based on~(\ref{eq:aleph}),~(\ref{eq:chi}) and~(\ref{eq:varrho}); \;
			Obtain the reward $r^{\scriptscriptstyle(1)}$ based on $\mathbf{a}^{\scriptscriptstyle(1)}$ and $\left\{\bm{B}^{\star},\left\{\eta_n^{\star}, \bm{w}_n^{\star}\right\}_{n\in \mathcal{N}}\right\}$ optimized by Alg.~\ref{alg:1}; \;
			Update $\theta_{\text{p}}$ and $\phi_{\text{p}}$ based on~(\ref{eq:La}) and~(\ref{eq:Lc}), respectively.}
		\If{$\mathcal{D}$ \rm{is filled with samples}}{
			Randomly select a transition batch from $\mathcal{D}$; \;
			Update soft Q-function parameters with~(\ref{eq:Q}); \;
			Update policy network parameter with~(\ref{eq:pi});\;
			Update target soft Q-function parameters with~(\ref{eq:theta1}).\;
	}}
\end{algorithm}

\subsubsection{MDP for Sensor Scheduling} \label{sec:MDPS}
This MDP focuses on the sensor scheduling, where the pointer network receives a sequence of encoded features representing the UAV trajectory and sensor location information, and outputs a scheduling sequence. In the Actor-Critic framework, these encoded features serve as the state for policy evaluation, while the reward reflects the system's performance resulting from the scheduling strategy. Formally, this MDP can be defined as follows.
\begin{itemize}
	\item {\em State Space $\mathcal{S}^{\scriptscriptstyle(1)}$}: The state $\mathbf{s}^{\scriptscriptstyle(1)} \in \mathcal{S}^{\scriptscriptstyle(1)}$ is denoted as
	\begin{equation}
		\mathbf{s}^{\scriptscriptstyle(1)} = \left\{ \left[\bm{q}^\text{u}_n\right]_{n\in\mathcal{N}}; \left[\bm{q}_{ji}\right]_{j\in\mathcal{J},i\in\mathcal{L}_j}  \right\},
	\end{equation}
	which has a cardinality of $3\left(N+L\right)$.
	
	\item {\em Action Space $ \mathcal{A}^{\scriptscriptstyle(1)} $}: The action $\mathbf{a}^{\scriptscriptstyle(1)} \in \mathcal{A}^{\scriptscriptstyle(1)}$ is defined as
	\begin{equation}
	\mathbf{a}^{\scriptscriptstyle(1)}=\left[\textstyle\sum_{n=1}^{N}n\cdot c_{j,n}\cdot a_{ji,n}\right]_{j\in\mathcal{J},i\in\mathcal{L}_j},
	\end{equation}
	where each element of $\mathbf{a}^{\scriptscriptstyle(1)}$ represents the scheduled transmission slot of the corresponding sensor. In the pointer network, each sensor's scheduled slot can be generated sequentially by a mask mechanism and a pointer generator, ensuring that the scheduling variables satisfy the constraints in~(\ref{eq:c0}) and~(\ref{eq:a0}) in the problem~(\ref{eq:p0}).
	
	\item {\em Reward $r^{\scriptscriptstyle(1)}(\mathbf{s}^{\scriptscriptstyle(1)}, \mathbf{a}^{\scriptscriptstyle(1)})$}: The reward quantifies the effectiveness of the scheduling strategy chosen by the agent in a given state, reflecting its influence on the objective function in~(\ref{eq:p0}) under a fixed UAV trajectory. Specifically, it captures the impact on both the MSE and the transmission power, and is formally defined as follows.
	\begin{equation}
		r^{\scriptscriptstyle(1)}(\mathbf{s}^{\scriptscriptstyle(1)}, \mathbf{a}^{\scriptscriptstyle(1)})= \kappa_1\textstyle\sum\limits_{j\in \mathcal{J}}\text{MSE}_j +\kappa_2 E^{\text{tran}},
	\end{equation}
	where $\textstyle\sum\nolimits_{j\in \mathcal{J}}\text{MSE}_j$ and $E^{\text{tran}}$ are computed based on the inner-layer AirComp transceiver strategy optimized by Alg.~\ref{alg:1} and the fixed trajectory.
\end{itemize}

\subsubsection{MDP for UAV Trajectory} \label{sec:MDPT}
This MDP focuses on the UAV trajectory planning and can be defined as follows.

\begin{itemize}
	\item {\em State Space $\mathcal{S}^{\scriptscriptstyle(2)}$}: The state $\mathbf{s}_{n}^{\scriptscriptstyle(2)} \in \mathcal{S}^{\scriptscriptstyle(2)}$, $n\in\mathcal{N}$, corresponds to the system environment state at slot $n$, denoted as
	\begin{equation}
		\begin{split}
			\mathbf{s}_{n}^{\scriptscriptstyle(2)} =\:&\left\{ \left\{c_{j,n}\right\}_{\forall j\in\mathcal{J}}; \left\{a_{ji,n}\right\}_{\forall j\in\mathcal{J}, \forall i\in\mathcal{L}_j}; \right.\\
			&\left.\bm{q}^\text{u}_n; v_n; \bm{\text{CMSE}}_n; E_n^{\text{tran}}; E_n^{\text{prop}} \right\},
		\end{split}
	\end{equation}
	which has a cardinality equal to $(2J+L+6)$. Here, $\bm{\text{CMSE}}_n$, $E_n^{\text{tran}}$, and $E_n^{\text{prop}}$ denote the vector of cumulative computational errors for all clusters, the cumulative transmission energy consumption, and cumulative propulsion energy consumption in slot $n$, respectively.
	
	\item {\em Action Space $ \mathcal{A}^{\scriptscriptstyle(2)} $}: The action $\mathbf{a}_{n}^{\scriptscriptstyle(2)} \in \mathcal{A}^{\scriptscriptstyle(2)}$, $n\in\mathcal{N}$ is defined by a 3-dimensional vector $\mathbf{a}_{n}^{\scriptscriptstyle(2)}=\left\{\psi^v_n, \psi^h_n, \psi^a_n\right\}$, where $\psi^v_n$ and $\psi^h_n$ represent the polar and azimuthal angles of the flight direction, respectively, while $\psi^a_n$ denotes the UAV acceleration.
	
	\item {\em Reward $r^{\scriptscriptstyle(2)}(\mathbf{s}^{\scriptscriptstyle(2)}_{n}, \mathbf{a}_{n}^{\scriptscriptstyle(2)})$}: The reward measures the effect of the UAV flight control taken by an agent for a given state, which further guides the agent to find the best trajectory. To satisfy various requirements such as AirComp accuracy, energy consumption, and the specified UAV destination, the reward function incorporates two components derived from the objective function of the problem in~(\ref{eq:p0}) and the UAV destination constraint, denoted by $r_{1,n}$ and $r_{2,n}$, respectively. Specifically, $r_{1,n}$ is given as
	\begin{equation}
		\begin{split}
			r_{1,n} =&\:\kappa_1\textstyle\sum\limits_{j \in \mathcal{J}}c_{j,n}\bigg(\textstyle\sum\limits_{i \in \mathcal{L}_j}a_{ji,n}\left| \eta_n\bm{w}^{\text{H}}_n\bm{g}_{ji,n}b_{ji,n} - 1\right| ^2\bigg.\\
			&\bigg. +\eta_n^2\bm{w}^{\text{H}}_n \bm{w}_n\sigma^2\bigg) \!+ \!\kappa_2\!\!\textstyle\sum\limits_{j \in \mathcal{J}}\!c_{j,n}\!\!\textstyle\sum\limits_{i \in \mathcal{L}_j}\!\!a_{ji,n}|b_{ji,n}|^2\\
			&+\kappa_2\cdot\varsigma P(v_n),\qquad
		\end{split}
	\end{equation}
	where the first term denotes the instantaneous computation error, and the second and third terms are the normalized transmission and flying energy consumptions in the current slot. The normalization is applied to balance their numerical scales and ensure comparable influence. Meanwhile,
	\begin{equation}
		r_{2,n}= \left\{\!\!\!\!\!\!\!\begin{array}{rcl}
			&r_f,~\text{if}~n=N~\text{and}~\sqrt{\left\|\bm{q}^{\text{u}}_n-\bm{q}_f\right\|^2} \leq v_{\max}\delta,\\
			&\!\!\!\!\!\!\!\!\!\!\!\!\!\!\!\!\!-\kappa_3 \cdot n \times \sqrt{\left\|\bm{q}^{\text{u}}_n-\bm{q}_f\right\|^2},~\text{otherwise}, \end{array}\right.
	\end{equation}
	where $r_f$ represents the arrival reward and $\kappa_3$ is a proportional coefficient. Therefore, the total reward is $r^{\scriptscriptstyle(2)}(\mathbf{s}^{\scriptscriptstyle(2)}_{n}, \mathbf{a}_{n}^{\scriptscriptstyle(2)})=r_{1,n}+r_{2,n}$.
\end{itemize}

\subsection{Pointer Actor-Critic Learning for Scheduling} \label{sec:PN-AC}
As illustrated in Fig.~\ref{fig:pn-sac}, the proposed pointer actor-critic framework is composed of two cooperative subnetworks, namely the pointer network and the critic network. The pointer network functions as the actor to produce the scheduling action $\mathbf{a}^{\scriptscriptstyle(1)} \in \mathcal{A}^{\scriptscriptstyle(1)}$, determining which sensors are activated for data transmission. The critic network, designed as an encoder-based feedforward architecture employing 1D pointwise convolutions, estimates the state–action value function $Q_{\theta_{\text{pn}}}(\cdot)$, where $\theta_{\text{pn}}$ denotes the trainable parameters of the critic. This value function serves to evaluate the benefit of the current action in terms of its contribution to improving the overall energy efficiency under the given system state. Moreover, to reduce the model complexity and enhance training efficiency, the two subnetworks share the encoder parameters of the Pointer Network during training.

The proposed pointer network follows a sequence to sequence paradigm, which is composed of an encoder, a decoder, and a pointer generation unit~\cite{pn}. Different from conventional implementations based on recurrent cells, both the encoder and decoder in our design are built upon convolutional neural network blocks to enhance parallel processing and feature abstraction efficiency. The input of the pointer network corresponds to the state representation $\mathbf{s}^{\scriptscriptstyle(1)} \in \mathcal{S}^{\scriptscriptstyle(1)}$, formulated as the concatenated sequence $\left[\left[\bm{q}^{\text{u}}_n\right]_{n\in\mathcal{N}\cup \{0\}}; \left[\bm{q}_{ji}\right]_{j\in\mathcal{J}, i\in\mathcal{L}_j}\right]$, where each component encapsulates the position coordinates of UAV and sensors. The network outputs a set of pointers $\mathbf{a}^{\scriptscriptstyle(1)}=\left[\textstyle\sum_{n=1}^{N}n\cdot c_{j,n}\cdot a_{ji,n}\right]_{j\in\mathcal{J}, i\in\mathcal{L}_j}$, representing the scheduling indices of sensors within each cluster. During the forward propagation, the encoder extracts compact feature representations from the input sequence, while the decoder, together with the pointer generator, produces the attention-driven pointers. Let the encoder feature maps and the decoder embeddings for the $i$-th sensor in cluster $j$ be denoted by $\overline{\bm{Q}}^{\text{u}}_{ji} = \left[\overline{\bm{q}}^{\text{u}}_0, \cdots, \overline{\bm{q}}^{\text{u}}_N\right]$ and $\overline{\bm{q}}_{ji}$, respectively. The correlation between the encoded UAV trajectory and the current decoding state is obtained through the attention operation, which is expressed as
\begin{equation}\label{eq:aleph}
\aleph_{ji}= \text{softmax}\left(f_{\text{A}}\left(\overline{\bm{Q}}^{\text{u}}_{ji}, \overline{\bm{q}}_{ji}\right)\right) \left(\overline{\bm{Q}}^{\text{u}}_{ji}\right)^{\text{T}},~j\in\mathcal{J}, i\in\mathcal{L}_j.
\end{equation}
Here, $f_{\text{A}}(\cdot)$ is the attention module that generates an aggregated feature $\aleph_{ji}$ by emphasizing the most relevant UAV waypoints. This process enables the decoder to focus on the UAV position and the corresponding time slot that contributes most to the current scheduling decision. Based on the derived context, the pointer generator evaluates the relative importance of each candidate by
\begin{equation} \label{eq:chi}
\chi_{ji} = \mathbf{v}^\text{T} \tanh\left(\mathbf{M} \left[\:\overline{\bm{Q}}^{\text{u}}_{ji,n}; \aleph_{ji}\right]\right),~j\in\mathcal{J}, i\in\mathcal{L}_j,
\end{equation}
where $\mathbf{v}$ and $\mathbf{M}$ are trainable parameters. The nonlinear projection fuses the attention context with the encoded features and produces a relevance score $\mu_{ji}$ for each candidate sensor, which can then be normalized to form a probability distribution, as
\begin{equation} \label{eq:varrho}
\varrho_{ji} = \text{softmax}(\mu_{ji}),~j\in\mathcal{J}, i\in\mathcal{L}_j,
\end{equation}
where $\varrho_{ji}$ represents the probability assigned to each candidate. Finally, the $\arg\max(\cdot)$ operator is employed as the pointer selection function to determine the scheduling decision by choosing the index corresponding to the highest probability.

In the Actor-Critic algorithm, the critic network estimates the state-value function $V_{\theta_{\text{p}}}(\cdot)$, which represents the expected cumulative reward starting from the current state. $V_{\theta_{\text{p}}}(\cdot)$ serves as a baseline for evaluating the relative quality of actions to reduce the variance of policy updates during training. To quantify the relative value of a selected action, the advantage function is introduced and defined as 
\begin{equation}
	A^{\pi_{\phi_\text{p}}}(\mathbf{s}^{\scriptscriptstyle(1)},\mathbf{a}^{\scriptscriptstyle(1)}) = r^{\scriptscriptstyle(1)} + V_{\theta_{\text{p}}}(\acute{\mathbf{s}}^{\scriptscriptstyle(1)}) - V_{\theta_{\text{p}}}(\mathbf{s}^{\scriptscriptstyle(1)}),
\end{equation}
capturing the gain of an action relative to the expected value of the current state and providing a guidance signal for policy improvement. During the training phase, the actor and critic networks update their parameters via gradient descent and by minimizing the mean squared error, respectively. Specifically, the pointer network loss and the critic loss can be computed as follows, respectively,
\begin{equation} \label{eq:La}
	L_{\text{a}}(\phi_{\text{p}}) = -\log \pi_{\phi_{\text{p}}}(\mathbf{s}^{\scriptscriptstyle(1)}, \mathbf{a}^{\scriptscriptstyle(1)}) \left(r^{\scriptscriptstyle(1)} - V_{\theta_{\text{p}}}(\mathbf{s}^{\scriptscriptstyle(1)}) \right),
\end{equation}
\begin{equation} \label{eq:Lc}
	L_{\text{c}}(\theta_{\text{p}}) = \left( V_{\theta_{\text{p}}}(\mathbf{s}^{\scriptscriptstyle(1)}) - r^{\scriptscriptstyle(1)} \right)^2.
\end{equation}
The training procedures of the pointer actor-critic network for scheduling are summarized as line 12 to line 17 in Alg.~\ref{alg:2}.

\subsection{Soft Actor-Critic Algorithm for Trajectory} \label{sec:SAC-TO}
As shown in Fig.~\ref{fig:pn-sac}, the proposed SAC-based algorithm consists of a policy network, two soft Q-function networks, two target soft Q-function networks, and a replay buffer. Unlike traditional reinforcement learning algorithms, the SAC algorithm incorporates the concept of maximum policy entropy in the objective function, as formulated as follows
\begin{equation}
	\begin{split}
		&\max_{\pi}\sum_{n=1}^N~\mathbb{E}_{\left(\mathbf{s}_{n}^{\scriptscriptstyle(2)}, \mathbf{a}_{n}^{\scriptscriptstyle(2)}\right)\sim\zeta_{\pi}}\left[\gamma^{n-1}r^{\scriptscriptstyle(2)}(\mathbf{s}_{n}^{\scriptscriptstyle(2)}, \mathbf{a}_{n}^{\scriptscriptstyle(2)}) \right. \\
		&\left.-\beta\log\left(\pi(\mathbf{a}_{n}^{\scriptscriptstyle(2)}|\mathbf{s}_{n}^{\scriptscriptstyle(2)})\right)\right],
	\end{split}
\end{equation}
where $\zeta_{\pi}$ is the state-action marginal distribution following $\pi$, $\gamma\in(0,1)$ is a discount factor, and $\beta$ is the temperature parameter that determines the weight of the entropy versus the reward. This design considers both the immediate reward of the policy and its randomness, enhancing the stability and robustness of the algorithm, and achieving a balance between exploration and exploitation~\cite{sac}. Due to the incorporation of the entropy, the state-value function in SAC is given as
\begin{equation}
	\mathbf{v}_{\pi}(\mathbf{s}_{n}^{\scriptscriptstyle(2)}) = \mathbb{E}_{\mathbf{a}_{n}^{\scriptscriptstyle(2)}\sim{\pi}}\left[Q_{\pi}(\mathbf{s}_{n}^{\scriptscriptstyle(2)}, \mathbf{a}_{n}^{\scriptscriptstyle(2)})-\beta\log\left(\pi(\mathbf{a}_{n}^{\scriptscriptstyle(2)}|\mathbf{s}_{n}^{\scriptscriptstyle(2)})\right)\right],
\end{equation}
where $Q_{\pi}(\mathbf{s}_{n}^{\scriptscriptstyle(2)}, \mathbf{a}_{n}^{\scriptscriptstyle(2)})$ is the state-action value. Additionally, the introduction of double Q-function networks aims to mitigate the overestimation bias in Q-value estimation~\cite{dq}. The policy network is modeled as a probabilistic distribution over actions, typically a Gaussian whose mean and covariance are produced by a neural network. The raw action is sampled from the distribution, and mapped to the actual action space to obtain the executed action.

During the training phase of the SAC-based algorithm, two soft Q-function networks, two target soft Q-function networks, and a policy network are updated, with their corresponding parameters denoted as $\left\{\theta_1, \theta_2, \theta'_1, \theta'_2, \phi\right\} $ . When training, a batch of $D$ transition tuples $(\mathbf{s}_{n}^{\scriptscriptstyle(2)}, \mathbf{a}_{n}^{\scriptscriptstyle(2)}, r^{\scriptscriptstyle(2)}(\mathbf{s}_{n}^{\scriptscriptstyle(2)}, \mathbf{a}_{n}^{\scriptscriptstyle(2)}),\mathbf{s}_{n+1}^{\scriptscriptstyle(2)})$ are randomly selected from the replay buffer $\mathcal{D}$ for learning. The parameters of soft Q-function network $\theta_t,t=1,2,$ are updated by minimizing the soft Bellman residual defined as
\begin{equation} \label{eq:Q}
\begin{aligned}
	&J_{Q}\left(\theta_{t}\right) =\mathbb{E}_{\left(\mathbf{s}_{n}^{\scriptscriptstyle(2)}, \mathbf{a}_{n}^{\scriptscriptstyle(2)}\right) \sim \mathcal{D}}\!\Bigg[\frac { 1 } { 2 } \bigg(Q_{\theta_{t}}\!\left(\mathbf{s}_{n}^{\scriptscriptstyle(2)}, \mathbf{a}_{n}^{\scriptscriptstyle(2)}\right)-\!\bigg(\!r^{\scriptscriptstyle(2)}\!\left(\mathbf{s}_{n}^{\scriptscriptstyle(2)}, \mathbf{a}_{n}^{\scriptscriptstyle(2)}\right)\bigg.\bigg.\Bigg. \\
	& \Bigg.\bigg.\bigg.+\!\gamma\!\left(\min_{t=1,2} Q_{\theta'_{t}}\left(\mathbf{s}_{n+1}^{\scriptscriptstyle(2)}, \mathbf{a}_{n+1}^{\scriptscriptstyle(2)}\right)\!-\!\beta \log \pi_{\phi}\left(\mathbf{a}_{n+1}^{\scriptscriptstyle(2)}\!\!\mid\! \mathbf{s}_{n+1}^{\scriptscriptstyle(2)}\right)\!\!\right)\!\!\!\bigg)\!\!\bigg)^{2}\Bigg].
\end{aligned}
\end{equation}
Meanwhile, the parameter of policy function $\phi$ is updated by
\begin{table}[h]\centering
	\caption{Simulation Parameters}
	\label{table}
	\renewcommand{\arraystretch}{1.5}
	\begin{tabular}{c|c|c|c}
		\hline
		\textbf{Parameter} & \textbf{Value} & \textbf{Parameter} & \textbf{Value} \\
		\hline
		$\left[H_{\min}, H_{\max}\right]$ & [100, 300]\:m & $(\varepsilon_1, \varepsilon_2)$ & (9.613, 0.158)\\
		$v_{\max}$ & 35\:m/s & $(\mu_{\mathsf{LoS}}, \mu_{\mathsf{NLoS}})$ & (1\:dB, 20\:dB)\\
		$p_{\max}$ & 0.5\:W  & $\Gamma$ & 0.001\\
		$K$        & 4       & $\alpha_{q}$ & 1$\times$10{\textsuperscript{-4}}\\
		$T$        & 60\:s   & $\alpha_{\pi}$ & 1$\times$10{\textsuperscript{-4}}\\
		$\delta$   & 1\:s    & $\gamma$ & 0.90\\
		$\sigma^2$ & -140\:dBm & $D$ & 64\\
		$f_c$      & 2\:GHz  & $\left| \mathcal{D} \right|$ & 10{\textsuperscript{6}}\\
		\hline        
	\end{tabular} 
\end{table}

\begin{equation} \label{eq:pi}
\begin{aligned}
	J_{\pi}(\phi)=~&\mathbb{E}_{\mathbf{s}_{n}^{\scriptscriptstyle(2)} \sim \mathcal{D}, \epsilon_{n}}\Big[\beta \log \pi_{\phi}\left(f_{\phi}\left(\epsilon_{n} ; \mathbf{s}_{n}^{\scriptscriptstyle(2)}\right) \mid \mathbf{s}_{n}^{\scriptscriptstyle(2)}\right)\Big. \\
	& \Big.-\min_{t=1,2} Q_{\theta_{t}}\left(\mathbf{s}_{n}^{\scriptscriptstyle(2)}, f_{\phi}\left(\epsilon_{n} ; \mathbf{s}_{n}^{\scriptscriptstyle(2)}\right)\right)\Big],
\end{aligned}
\end{equation}
where $\epsilon_{n}$ is an independent noise vector and $\mathbf{a}_{n}^{\scriptscriptstyle(2)} = f_{\phi}\left(\epsilon_{n} ; \mathbf{s}_{n}^{\scriptscriptstyle(2)}\right)$. Then, the optimal $\beta$ can be obtained after solving for $Q^{\star}_{\theta_{i}}$ and $\pi^{\star}_\phi$:
\begin{equation} \label{eq:beta}
\beta^{\star}=\arg \min _{\beta} \mathbb{E}_{\mathbf{a}_{n}^{\scriptscriptstyle(2)} \sim \pi_\phi^{\star}}\left[-\beta \log \pi_{\phi}^{\star}\left(\mathbf{a}_{n}^{\scriptscriptstyle(2)} \mid \mathbf{s}_{n}^{\scriptscriptstyle(2)}; \beta\right)-\beta \overline{\mathcal{H}}\right],
\end{equation}
where $\overline{\mathcal{H}}=-\prod\limits_{\overline{m}=1}^{M+2}\left(\varpi_{\overline{m}}\right)$ is the target entropy based on the action space dimension, and $\varpi_{\overline{m}}$ is the space dimension of the $\overline{m}$-th action. Finally, based on the soft update method, the target soft Q-function network are updated as
\begin{equation} \label{eq:theta1}
  \theta'_{t} \gets \ell \theta_{t} + \left(1 - \ell\right)\theta'_{t}, \quad \ell\ll 1.
\end{equation}

\subsection{Pointer Network-Assisted SAC Learning} \label{sec:PNSAC}
As detailed in Alg.~\ref{alg:2}, the proposed pointer network-assisted SAC learning framework follows an offline training, online deployment paradigm, which effectively resolves the contradiction between the high computational overhead of the training phase and the stringent low-latency demands of online decision. Specifically, during the execution phase, before each SAC episode begins, the pointer network generates the scheduling strategy set $\left\{\widetilde{\bm{C}},\widetilde{\bm{A}}\right\}$ based on the latest UAV trajectory (for the initial episode, an initial scheduling strategy is used instead). The obtained $\left\{\widetilde{\bm{C}},\widetilde{\bm{A}}\right\}$ is combined with the optimized AirComp transceiver strategy by Alg.~\ref{alg:1} to determine the initial state $\mathbf{s}_1^{\scriptscriptstyle(2)}$. In each of the $N$ time slots, the SAC policy network $\pi_{\phi}$ outputs a UAV motion action $\mathbf{a}_{n}^{\scriptscriptstyle(2)}$ according to the current state $\mathbf{s}_{n}^{\scriptscriptstyle(2)}$. Then, executing $\mathbf{a}_{n}^{\scriptscriptstyle(2)}$ and Alg.~\ref{alg:1} yields the next state $\mathbf{s}_{n+1}^{\scriptscriptstyle(2)}$ and reward $r^{\scriptscriptstyle(2)}(\mathbf{s}^{\scriptscriptstyle(2)}_{n}, \mathbf{a}_{n}^{\scriptscriptstyle(2)})$, which are stored in the replay buffer. During the training phase, the pointer network is updated in intermittent manner. It leverages the UAV trajectories and user positions collected from recent training interval episodes as input data and combines them with the AirComp transceiver optimization by Alg.~\ref{alg:1} for parameter updating. Meanwhile, the SAC starts sampling and gradient updates once the replay buffer $\mathcal{D}$ is full, when newly generated samples replace the oldest ones.

\begin{figure}[t]
	\centering
	\includegraphics[width=8cm]{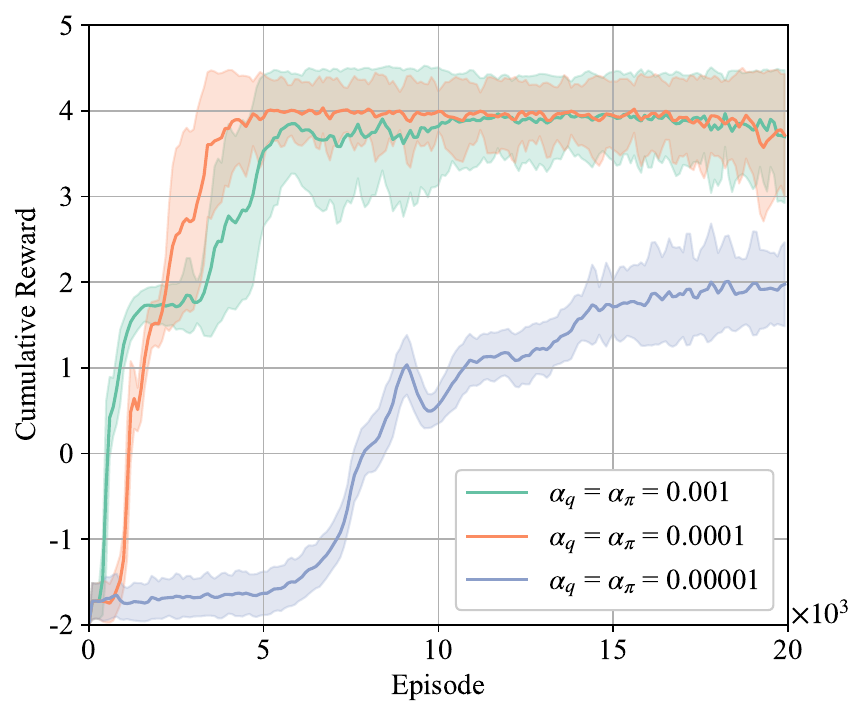}
	\vspace{-1.5mm}
	\caption{Cumulative reward vs. learning rates.}
	\label{fig:conver_r}
\end{figure}

\section{Simulation Results} \label{sec:sim}
In this section, we present the simulation results to show the performance. We consider an area of 1,000~m $ \times $ 1,000~m. There are 4 clusters, 36 sensors, and a UAV. The devices are randomly located within the area. The UAV propulsion energy related parameters are $P_0=79.86$, $W_{\text{tip}}=120\:\text{m/s}$, $P_i=88.63$, $v_0=4.03\:\text{m/s}$, $d_0=0.6$, $\rho=1.225\:\text{kg/m}^3$, $\omega=0.05$, $\Lambda=0.503\:\text{m}^2$. The weight for propulsion energy is $\varsigma=0.001$. The policy network and soft Q-function networks in the SAC each have two fully connected hidden layers, with 128 neurons in each layer. In the pointer actor-critic, the encoder is implemented as a convolutional layer with 128 neurons, while the critic network consists of 3 convolutional layers, each with 128 neurons. The learning rate and batch size of the pointer actor-critic are set to 0.0001 and 128 respectively. All experiments were run in Python 3.8 on a platform with an Intel i5-12600KF CPU, and an NVIDIA GeForce RTX 3070 GPU. The o parameters are summarized in Table~\ref{table}, used as defaults unless otherwise noted.

\subsection{Convergence of pointer actor-critic learning}
In Fig.~\ref{fig:conver_r}, we evaluate the convergence behavior of the pointer actor-critic algorithm under varying learning rates. Specifically, it is demonstrated that all curves eventually converge, thus validating the inherent stability of the proposed algorithm. The convergence trends of the curves, coupled with the variance represented by the shaded regions, illustrate that the learning rate exerts a significant impact on both the convergence speed and stability of the algorithm. A relatively large learning rate accelerates the initial convergence process yet introduces considerable fluctuations, which is presumably attributed to the overshooting of optimal parameters across the error surface. Conversely, a relatively small learning rate leads to sluggish convergence while yielding smoother trajectories, owing to the more gradual updates of model parameters. This underscores the critical importance of selecting an optimal learning rate to strike a balance between training efficiency and convergence stability.

\begin{figure}[t]
	\centering
	\includegraphics[width=7.4cm]{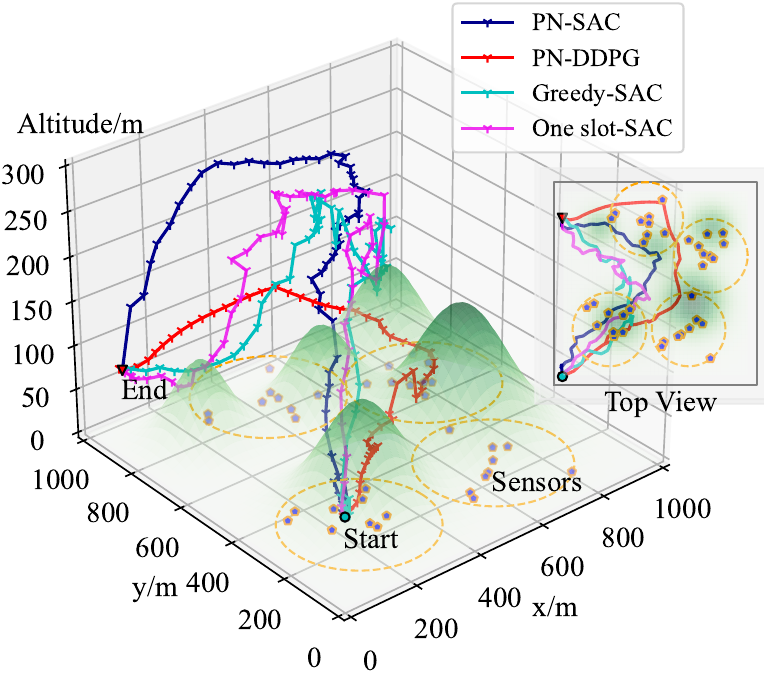}
	\vspace{-1.5mm}
	\caption{Trajectories vs. algorithms.}
	\vspace{-3.5mm}
	\label{fig:traj_diff_alg}
\end{figure}
\begin{figure}[t]
	\centering
	\includegraphics[width=7.4cm]{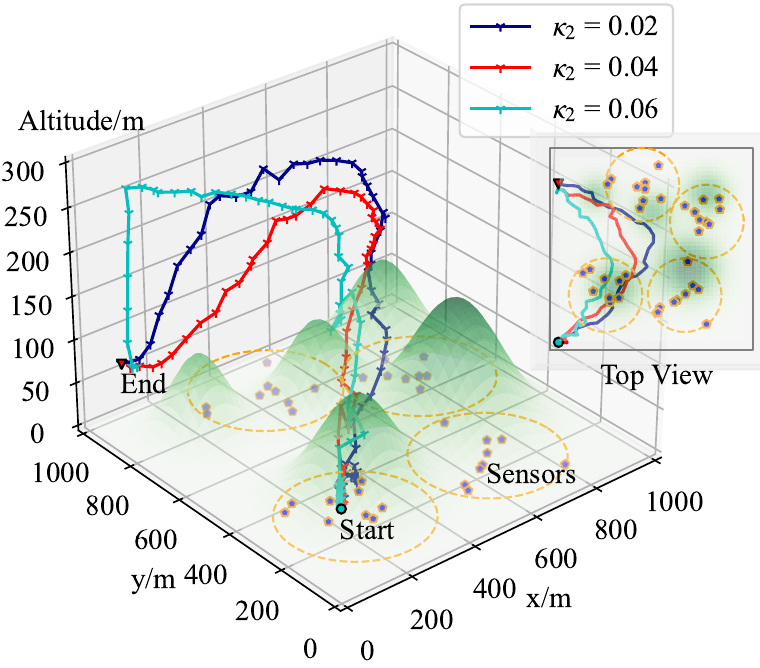}
	\vspace{-1.5mm}
	\caption{Trajectories vs. energy weights.}
	\vspace{-3.5mm}
	\label{fig:traj_diff_param}
\end{figure}
\begin{figure}[t]
	\centering
	\includegraphics[width=8cm]{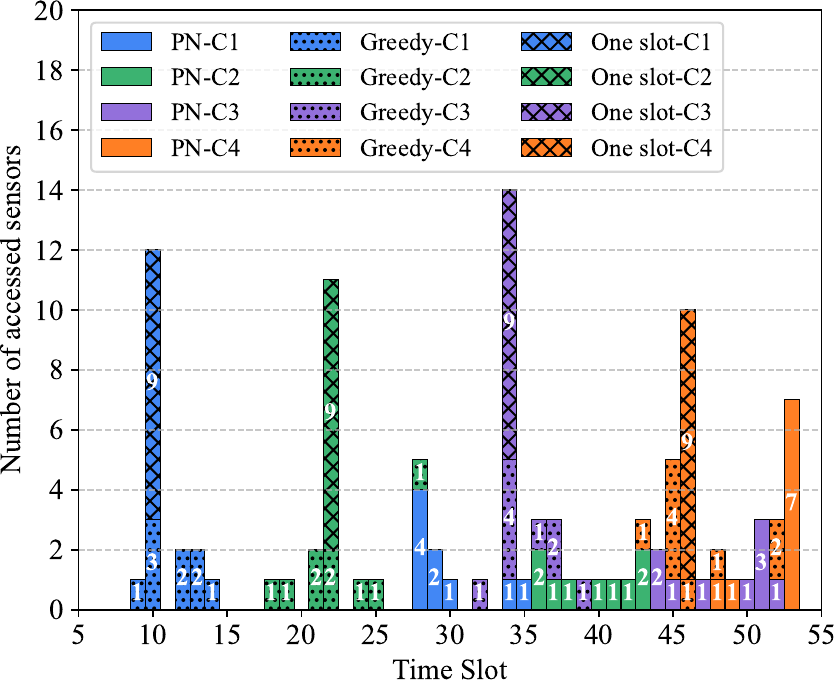}
	\vspace{-1.5mm}
	\caption{Sersor scheduling scheme over the timeline.}
	\vspace{-3.5mm}
	\label{fig:sched_slot}
\end{figure}

In Fig.~\ref{fig:traj_diff_alg}, we compare the UAV trajectories optimized by four methods, including the proposed pointer network-assisted SAC (PN-SAC), the pointer network-assisted DDPG (PN-DDPG), the greedy scheduling-based SAC (Greedy-SAC), and the single-slot scheduling-based SAC (One-slot SAC). The PN-DDPG is used to compare the UAV trajectory optimization performance of SAC in the PN-SAC framework, while the Greedy-SAC and One-slot SAC are employed to evaluate the impact of the pointer network on scheduling strategies. It can be observed that the pointer network-based methods exhibit more adaptive trajectory patterns, where the UAV dynamically adjusts both altitude and horizontal positions to match the spatial distribution of sensors. This adaptability arises from the multi-slot scheduling capability of the pointer network, which allows the UAV to adapt to channel fading effects. In contrast, the One-slot SAC is constrained by its single-slot transmission mechanism and thus lacks the ability to optimize trajectories over multiple time slots, whereas the Greedy-SAC prioritizes immediate rewards at the expense of maintaining favorable long-term channel conditions. Meanwhile, the PN-SAC demonstrates greater utilization of the UAV's flight altitude compared to the PN-DDPG. By flexibly adjusting its altitude, PN-SAC increases the probability of LoS links, thereby reducing both the MSE and transmission energy consumption. This improvement is attributed to the enhanced exploration capability of SAC, enabled by the entropy term introduced in its objective function. Therefore, integrating the pointer network with SAC facilitates more robust trajectory planning, effectively mitigating computational errors induced by channel mismatch in UAV-assisted AirComp scenarios.

In Fig.~\ref{fig:traj_diff_param}, we present the variations of UAV trajectories generated by the proposed algorithm under different energy consumption weights $\kappa_2$. It can be observed that as $\kappa_2$ increases, the UAV tends to reduce the flight distance during the entire task and diminish the tendency to move closer to the sensor distribution areas. This is because a larger $\kappa_2$ strengthens the algorithm's preference for reducing system energy consumption, which drives the UAV to shorten its flight path even at the cost of sacrificing the computational accuracy of AirComp. The trajectory results show that adjusting $\kappa_2$ enables our algorithm to dynamically balance energy consumption and computational accuracy, which conforms to the joint optimization goal of minimizing both metrics.

In Fig.~\ref{fig:sched_slot}, we compare the scheduling strategies optimized by the greedy algorithm, the one-slot method and the pointer actor-critic algorithm. C1 to C4 correspond to four sensor clusters. By combining the UAV trajectory shown in Fig.~\ref{fig:traj_diff_alg}, it can be observed that the pointer actor-critic algorithm accesses sensors within a specific cluster during the time slot when the UAV flies near that cluster and schedules sensors based on instantaneous channel states. In contrast, the greedy algorithm determines scheduling solely according to instantaneous system states, thereby prioritizing short-term aggregation accuracy at the expense of long-term system performance. For the one-slot strategy, since the simultaneous transmission of sensors within the same cluster, it lacks adaptability to dynamic channel conditions.

\subsection{Performance Comparison}
\begin{figure}[t]
	\centering
	\includegraphics[width=8cm]{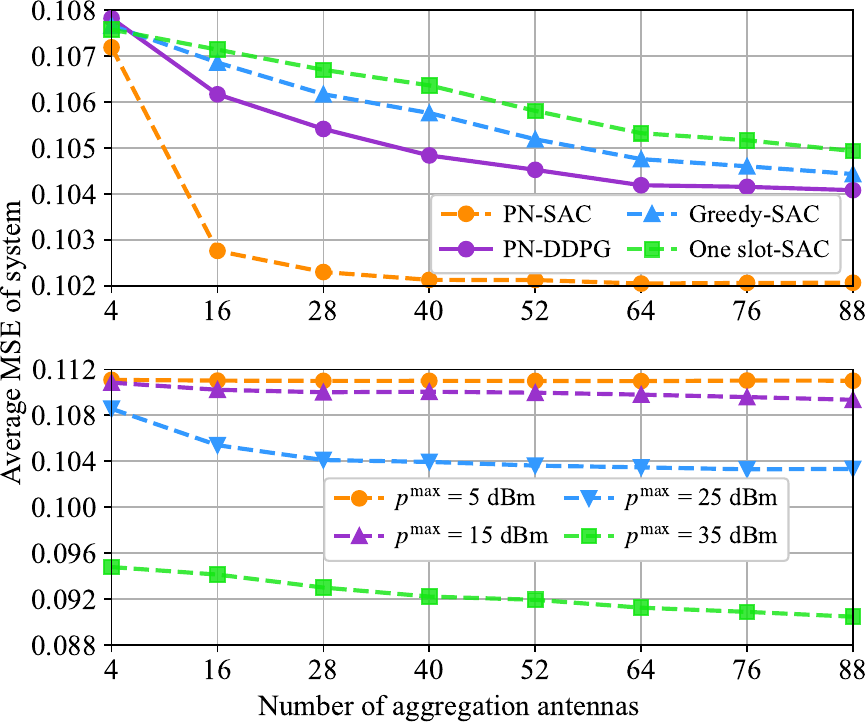}
	\vspace{-1.5mm}
	\caption{The system MSE vs. the number of UAV antennas.}
	\vspace{-7mm}
	\label{fig:5_diff_alg-pnsac_atenna}
\end{figure}

In the upper subfigure of Fig.~\ref{fig:5_diff_alg-pnsac_atenna}, we compare the system MSE achieved by four algorithms under different numbers of aggregation antennas. It can be observed that the system MSE for all algorithms decreases as the number of antennas increases, which is attributed to the enhanced beamforming gain that strengthens signal alignment and improves signal transmission quality. Owing to the joint optimization of precoding, beamforming, and multi-time-slot scheduling, PN-SAC achieves lower system MSE than the other algorithms, with PN-DDPG closely behind. Furthermore, Greedy-SAC and One-slot-SAC exhibit inferior performance because their scheduling strategies fail to adapt to dynamic and complex channel conditions. These results demonstrate that integrating the pointer network into SAC fully exploits antenna diversity and channel-aware scheduling to improve the performance.

In the lower subfigure of Fig.~\ref{fig:5_diff_alg-pnsac_atenna}, we analyze the system MSE performance of the proposed PN-SAC algorithm under different upper limits of sensor transmission power and varying numbers of aggregation antennas. It can be observed that a higher transmission power limit leads to a lower MSE, as stronger signals help mitigate fading effects. Increasing the number of antennas further amplifies this improvement through beamforming gains. It is worth noting that with 88 antennas, the lowest MSE is achieved when the maximum transmission power is 35~dBm. These results confirm that transmission power and antenna diversity can synergistically enhance signal quality in considered system.

\begin{figure}[t]
	\centering
	\includegraphics[width=8cm]{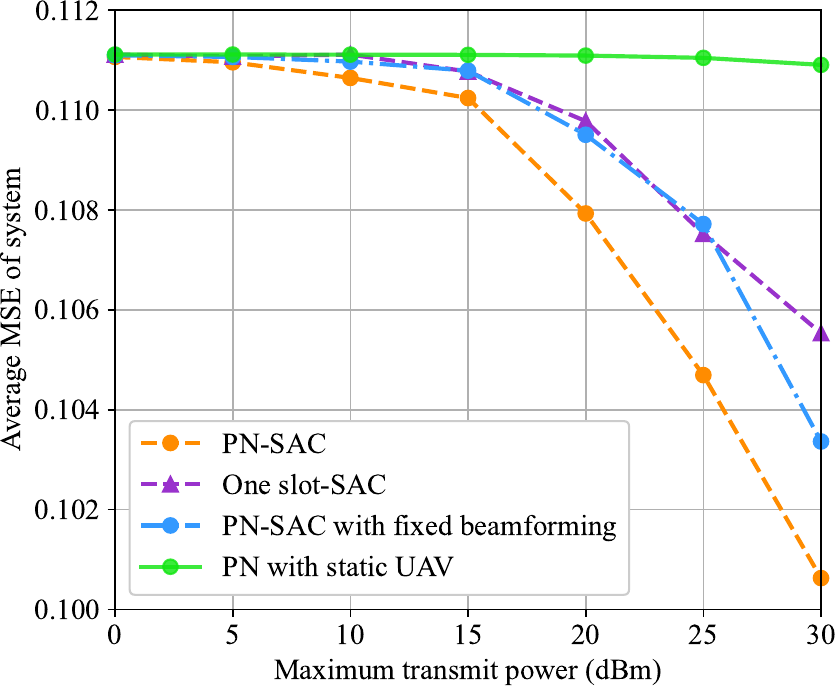}
	\vspace{-1.mm}
	\caption{The system MSE versus the power budgets under different methods.}
	\label{fig:diff_pnsac_power}
\end{figure}

In Fig.~\ref{fig:diff_pnsac_power}, we compare the system MSE performance of the PN-SAC, One-slot SAC, PN-SAC with fixed beamforming, and the static UAV method under different upper limits of sensor transmission power. As expected, PN-SAC consistently achieves the lowest MSE, owing to its dynamic trajectory design and adaptive scheduling, which effectively exploit power gains. In contrast, the static UAV method and One-slot SAC exhibit poorer performance because their ability to cope with dynamic channel variations is limited. Moreover, PN-SAC with fixed beamforming also performs worse, as it lacks one optimization dimension compared with the original PN-SAC. These results highlight the effectiveness of jointly optimizing trajectory, beamforming, and multi-time-slot scheduling.

\begin{figure}[t]
	\centering
	\includegraphics[width=8cm]{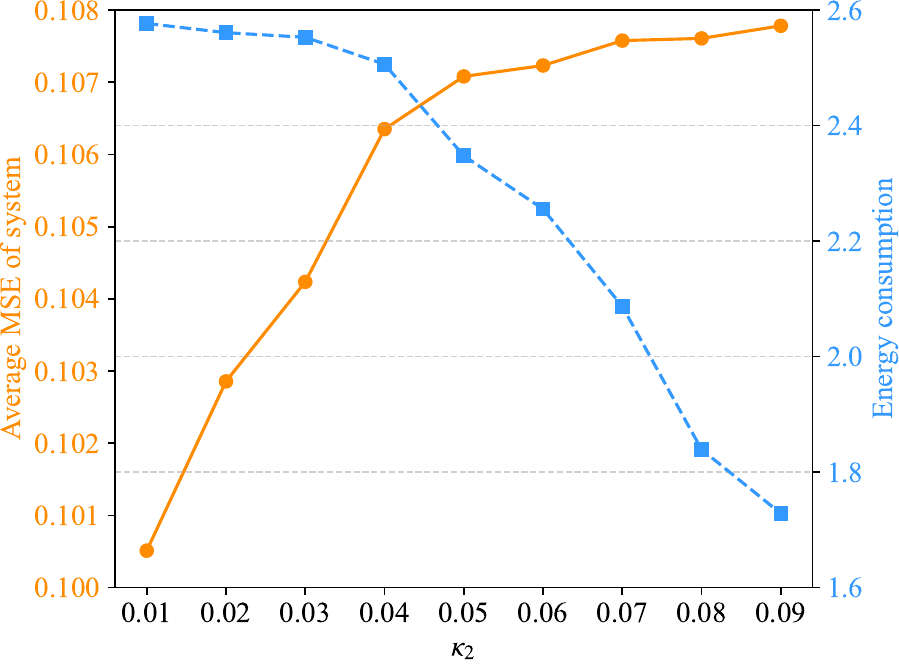}
	\vspace{-1.5mm}
	\caption{The system MSE and energy consumption vs. energy weights.}
	\vspace{-3.5mm}
	\label{fig:ec_mse}
\end{figure}

In Fig.~\ref{fig:ec_mse}, we conduct a sensitivity analysis with respect to the weighting coefficient $\kappa_2$ and present the variations of system energy consumption and MSE as $\kappa_2$ increases, where the system energy consumption corresponds to the energy term in the objective function in~(\ref{eq:ob0}). It can be observed that, as $\kappa_2$ increases, the system energy consumption decreases while the MSE rises, revealing a clear trade-off between the two metrics. This is because a larger $\kappa_2$ assigns higher priority to energy efficiency, while computational accuracy becomes relatively secondary in the optimization objective. This behavior is fully consistent with the dual-objective optimization goal and demonstrates that $\kappa_2$ serves as a balancing factor between minimizing UAV energy consumption and reducing computational error.
In addition, the curves also show that the sensitivity of the system performance to $\kappa_2$ is not uniform over the whole parameter range. When $\kappa_2$ becomes large, the reduction in system energy consumption gradually tends to saturate, whereas the MSE increases more noticeably, indicating that an overly aggressive emphasis on energy efficiency may lead to a disproportionate loss in computation accuracy. This observation further confirms that $\kappa_2$ is not merely a tuning parameter, but a meaningful design factor that determines the operating point of the proposed framework. Therefore, the negative correlation between system energy consumption and MSE not only validates the importance of adaptively selecting $\kappa_2$ according to scenario-specific priorities, but also shows that the adopted value of $\kappa_2$ corresponds to a representative trade-off point rather than an arbitrary choice.


\section{Conclusion} \label{sec:con}
In this paper, we propose a UAV-assisted time-slotted multi-cluster AirComp framework, which jointly optimizes sensor transmission, receive normalization factor, beamforming, sensor scheduling, and UAV trajectory. To tackle the problem, an alternating optimization approach is employed for the AirComp transceiving strategy in the inner layer. Additionally, a pointer actor-critic algorithm is introduced for the scheduling strategy under random channel fading, and a SAC algorithm is adopted for UAV trajectory planning. Numerical results demonstrate that the proposed pointer network-assisted SAC approach effectively mitigates distortion caused by channel misalignment within clusters, exhibits robustness to random fading, achieves lower aggregation error, and reduces UAV energy consumption compared with baseline methods.
\bibliographystyle{IEEEtran}
\bibliography{mainBib}

\end{document}